# Very Sparse Stable Random Projections, Estimators and Tail Bounds for Stable Random Projections

Ping Li
Department of Statistics
Stanford University
Stanford, CA 94305
pingli98@stanford.edu

June 15, 2018

## Abstract

The method of *stable random projections* [39, 41] is popular for data streaming computations, data mining, and machine learning. For example, in data streaming, stable random projections offer a unified, efficient, and elegant methodology for approximating the $l_\alpha$ norm of a single data stream, or the $l_\alpha$ distance between a pair of streams, for any $0 < \alpha \leq 2$. [18] and [20] applied stable random projections for approximating the Hamming norm and the max-dominance norm, respectively, using very small $\alpha$. Another application is to approximate all pairwise $l_\alpha$ distances in a data matrix to speed up clustering, classification, or kernel computations. Given that stable random projections have been successful in various applications, this paper will focus on three different aspects in improving the current practice of stable random projections.

Firstly, we propose *very sparse stable random projections* to significantly reduce the processing and storage cost, by replacing the $\alpha$-stable distribution with a mixture of a symmetric $\alpha$-Pareto distribution (with probability $\beta$, $0 < \beta \leq 1$) and a point mass at the origin (with a probability $1 - \beta$). This leads to a significant $\frac{1}{\beta}$-fold speedup for small $\beta$. We analyze the rate of convergence as a function of $\beta$, $\alpha$, and the data regularity conditions. For example, when $\alpha = 1$ and the data have bounded second moments, then if we choose $\beta = \frac{1}{D^{1/2}}$, the rate of convergence would be $O\left(D^{-1/2}\right)$, which is fast even for moderate $D$. Here $D$ is the data dimension. Some numerical evaluations are conducted, on synthetic data, Web crawl data, and gene expression microarray data.

Secondly, we provide an improved estimator for recovering the original $l_\alpha$ norm from the projected data. The standard estimator is based on the (absolute) sample median [39, 19], while we suggest using the geometric mean. The sample median estimator is difficult to analyze precisely or non-asymptotically. The geometric mean estimator we propose is strictly unbiased and is easier to study. Moreover, the geometric mean estimator is more accurate, especially non-asymptotically. When $\alpha \to 0+$ (as considered in [18, 19, 20]), even asymptotically, the geometric mean estimator is still about $27\%$ more accurate in terms of variances. In addition, we show that, when $\alpha = 0+$, the maximum likelihood estimator has a simple form and is considerably more accurate than both the geometric mean and sample median estimators.

Thirdly, we provide an explicit answer to the basic question of how many projections (samples) are needed for achieving some pre-specified level of accuracy. [39, 19] did not provide a criterion that can be used in practice. The geometric mean estimator we propose allows us to derive sharp tail bounds which can be expressed in exponential forms with constants explicitly given. From these tail bounds, an analog of the Johnson-Lindenstrauss (JL) Lemma for dimension reduction in $l_\alpha$ follows: It suffices to use $k = O\left(\frac{\log(n)}{\epsilon^2}\right)$ projections to guarantee that the $l_\alpha$ distance between any pair of data points (among $n$ data points) can be estimated within a $1 \pm \epsilon$ factor with high probability, using the proposed geometric mean estimator.

# 1 Introduction

Stable random projections[39, 41] can be applied at least in two types of applications: approximating all pairwise distances and data stream computations.

## 1.1 Computing All Pairwise Distances

We start with a "data matrix" $\mathbf{A} \in \mathbb{R}^{n \times D}$, with $n$ rows and $D$ columns ($n$ data points in $D$ dimensions). For example, $\mathbf{A}$ can be the *term-by-document* matrix at Web scale. Many applications require computing all pairwise distances of $\mathbf{A}$, the exact computation of which would cost $O(n^2 D)$, infeasible at Web scale. Broder and his colleagues [14, 12] developed various versions of the *min-wise* sketching algorithm[13, 16] to approximate all pairwise *resemblance* distances, for syntactic clustering of the AltaVista Web crawls and for removing duplicate pages.

We do not have to use the resemblance distance. Instead, we could use some $l_\alpha$ distance, $0 < \alpha \leq 2$. Given two data points $u_1$ and $u_2$ in $D$ dimensions, the individual $l_\alpha$ norms and the $l_\alpha$ distance are

$$\left(\sum_{i=1}^{D} |u_{1,i}|^\alpha\right)^{1/\alpha}, \quad \left(\sum_{i=1}^{D} |u_{2,i}|^\alpha\right)^{1/\alpha}, \quad d_{(\alpha)} = \left(\sum_{i=1}^{D} |u_{1,i} - u_{2,i}|^\alpha\right)^{1/\alpha}. \tag{1}$$

The idea of stable random projections[39, 41] is to multiply the original data matrix $\mathbf{A} \in \mathbb{R}^{n \times D}$ with a non-adaptive random projection matrix $\mathbf{R} \in \mathbb{R}^{D \times k}$ sampled i.i.d. from an $\alpha$-stable distribution[60], resulting in a projected matrix $\mathbf{B} = \mathbf{A}\mathbf{R} \in \mathbb{R}^{n \times k}$. If $k$ is fixed and small (we will show precisely how small $k$ can be.), then the cost for computing all-pairwise distances will be reduced from $O(n^2 D)$ to just $O(n^2 k + nDk)$, provided we can *estimate* the original $l_\alpha$ distances in $\mathbf{A}$ from $\mathbf{B}$.

Given two data points $u_1$, $u_2$ (two rows in $\mathbf{A}$), we denote the corresponding projected vectors by $v_1, v_2 \in \mathbb{R}^k$, i.e., $v_1 = \mathbf{R}^T u_1$, $v_2 = \mathbf{R}^T u_2$. We recommend the following estimator to reconstruct $d_{(\alpha)}$, the distance between $u_1$ and $u_2$:

$$\hat{d}_{(\alpha),gm} = \frac{\prod_{j=1}^{k} |v_{1,j} - v_{2,j}|^{1/k}}{\text{some correction factor}}, \tag{2}$$

which is the geometric mean with corrections. Alternatively, [39, 19] proposed using the (absolute) sample median:

$$\hat{d}_{(\alpha),me} = \frac{\text{median}\{|v_{1,j} - v_{2,j}|, j = 1, 2, ..., k\}}{\text{some correction factor}}, \tag{3}$$

which is a special case of sample quantile estimators[30, 31, 52]. The sample median estimator $\hat{d}_{(\alpha),me}$ is not as accurate, especially when $k$ is not too large. Moreover, the theoretical analysis on $\hat{d}_{(\alpha),me}$ is not as convenient, especially non-asymptotically.

We will show that using $\hat{d}_{(\alpha),gm}$, it suffices to choose $k = O\left(\frac{\log(n)}{\epsilon^2}\right)$ so that the $l_\alpha$ distance between any pair of data points in $\mathbf{A}$ can be estimated within a $1 \pm \epsilon$ factor with high probability.

## 1.2 Data Stream Computations

In data stream computations[38, 32, 39, 7, 18], stable random projections can be used at least for (A): approximating the $l_\alpha$ frequency moments for individual streams; (B): approximating the $l_\alpha$ differences between a pair of streams; (C): approximating the number of non-zero items (the Hamming norm) in a stream using very small $\alpha$[18, 19]. Here we only consider $0 < \alpha \leq 2$; but we should mention that $\alpha > 2$ is also sometimes computed[4].

Massive data streams are fundamental in many modern data processing applications. Data streams come from Internet routers, phone switches, atmospheric observations, sensor networks, highway traffic conditions, finance data, and more[38, 32, 39, 7, 18]. Unlike in the traditional databases, it is not common to store massive data streams; and hence the processing is often done "on the fly." For example, in some situations, we only need to "visually monitor" the data by observing the time history of certain summary statistics, e.g., sum, number of distinct items, or any $l_\alpha$ norm.

If we are only interested in the sums (or the range-sums), the DLT priority sampling algorithm would be ideal, especially for positive data[26, 27, 3]. Although the priority sampling algorithm had been proposed and implemented commercially for a few years, only recently [58] proved the non-asymptotic variance (bound) and the best constant.

Note that the sum of positive items is the $l_1$ norm. [39] described the procedure to use Cauchy (which is 1-stable) random projections for approximating the $l_1$ norms (or $l_1$ differences) of general data streams. For a data stream, $u_1$, which contains pairs $(i, u_{1,i})$, $i \in \{1, 2, ..., D\}$, [39] suggested the following steps (at least for the ideal version):

- Choose $k = O\left(\frac{1}{\epsilon^2}\right)$. Initialize $v_{1,j} = 0$, $j = 1, 2, ..., k$.
- Generate a matrix $\mathbf{R} \in \mathbb{R}^{D \times k}$, with entries $r_{ij}$ i.i.d. samples of standard Cauchy.
- For each new pair $(i, u_{1,i})$, modify $v_{1,j} = v_{1,j} + r_{ij} u_{1,i}$, for each $j = 1, 2, ..., k$.
- Return median$\{|v_{1,1}|, |v_{1,2}|, ..., |v_{1,k}|\}$ as the approximate $l_1$ norm of $u_1$.

[18, 19] extended the above procedure to general $0 < \alpha \leq 2$. [39, 18, 19] did not provide a practical criterion for choosing the sample size $k$.

## 1.3 Comparing Data Streams Using Hamming Norms

[18, 19] proposed approximating the Hamming norms of data streams using stable random projections with small $\alpha$, because the $l_\alpha$ norm raised to the $\alpha$th power approximates the Hamming norm well if $\alpha$ is sufficiently small. The Hamming norm gives the number of non-zero items present in a single stream; and it is also an important measure of (dis)similarity when applied to a pair of streams[18, 19]. Note that for static data, one could approximate the Hamming norms directly by applying 2-stable (i.e., normal) random projections on the binary-quantized (0/1) data. [18, 19] considered the dynamic setting in that the data may be subject to frequent additions/subtractions.

In order to well approximate the Hamming norm, [18, 19] let $0 < \alpha < \epsilon/\log(U)$, where $U$ is the largest item (in absolute values) in the stream(s). [18, 19] considered that, if an estimator, say $\hat{d}$, approximates the truth, $d$, within a factor of $1 \pm \epsilon$, then $\hat{d}^\alpha$ will be within $(1 \pm \epsilon)^\alpha$ factor of $d^\alpha$. Our concern is, because $\alpha$ is very small, $(1 \pm \epsilon)^\alpha \approx 1 \pm \epsilon \alpha$. If $\alpha = \epsilon$, then we will end up with a $1 \pm \epsilon^2$ factor instead of the usual $1 \pm \epsilon$ factor we like to have.

In this study, we will provide strictly unbiased, geometric mean types of estimators for both the $l_\alpha$ norm and the $l_\alpha$ norm raised to the $\alpha$th power, as well as their tail bounds. For the case $\alpha \to 0+$, we will compare in detail the geometric mean estimator with the sample median estimator, as well as the maximum likelihood estimator. Very interestingly, the maximum likelihood estimator in this case has a simple convenient form, whose variance is only about one half of the variance of regular normal ($l_2$) random projections. In other words, stable random projections with very small $\alpha$ would be indeed ideal for approximating the Hamming norms, not only in the dynamic settings but also preferable in static data.

## 1.4 Which Norm ($\alpha$) to Use?

$\alpha = 2$ is the most thoroughly studied case[59]. When $\alpha = 2$, we could directly estimate the original $l_2$ distances from the projected $l_2$ distances. The Johnson-Lindenstrauss (JL) Lemma says we only need $k = \left(\frac{\log(n)}{\epsilon^2}\right)$ (constants are well-known) so that the (squared) $l_2$ distance between any pair of data points can be estimated within a $1 \pm \epsilon$ factor with high probability. Many different versions of the JL Lemma have been proved[43, 35, 42, 5, 22, 39, 40, 1, 6, 2].

The case $\alpha = 1$ is also very often encountered in practice. However, it has been proved by [10, 45, 11] that one can not hope to develop an estimator that is a metric for dimension reduction in $l_1$ without incurring large errors.

Other norms are also possible. In data streaming computations, as $\alpha$ increases, the $l_\alpha$ distance attributes more significance to a large individual component; and therefore varying $\alpha$ provides a tunable mechanism[34]. This argument applies directly also in the machine learning content. As a concrete example, [15] proposed a family of non-Gaussian radial basis kernels for SVM in the form of $K(x, y) = \exp\left(-\rho \sum_i |x_i^a - y_i^a|^b\right)$, for data points $x$ and $y$. (Here $b$ is our $\alpha$.) [15] showed that $b = 0.5$ in some cases gave better results in histogram-based image classifications.

The $l_\alpha$ norm with $\alpha < 1$ is now well-understood to be a natural measure of sparsity[24, 25]. Of course, this is why [18, 19] approximate the Hamming norm with the $l_\alpha$ norm using small $\alpha$. [20] adopted the similar idea to approximate the max-dominance norm in data streams using very small $\alpha$.

## 1.5 Paper Organization

The paper is organized as follows. Section 2 briefly reviews stable random projections and summarizes our main results. We will then introduce *very sparse stable random projections* in Section 3. We will study in Section 4 the estimators and tail bounds for recovering both the $l_\alpha$ norm and the $l_\alpha$ norm raised to the $\alpha$th power. Section 5 compares the proposed geometric mean estimator with the sample median estimator, particularly for the case $\alpha \to 0+$. Section 5 also proposes a bias-corrected maximum likelihood estimator for the case $\alpha \to 0+$. Section 6 concludes the paper.

## 2 Review of Stable Random Projections and Summary of Main Results

A random variable $x$ is symmetric $\alpha$-stable if its characteristic function can be written as $\mathrm{E}\left(\exp\left(\sqrt{-1}xt\right)\right) = \exp\left(-d^\alpha|t|^\alpha\right)$, where $d > 0$ is the scale parameter. We write $x \sim S(\alpha, d)$, which in general does not have a closed-form density function except for $\alpha = 2$ (normal) or $\alpha = 1$ (Cauchy).

**The basic fact**: if $z_1, z_2, ..., z_D$, i.i.d. are $S(\alpha, 1)$, then for any constants (i.e., the original data) $g_1, g_2, ..., g_D$, we have $\sum_{i=1}^{D} g_i z_i \sim S\left(\alpha, \left(\sum_{i=1}^{D} |g_i|^\alpha\right)^{1/\alpha}\right)$. That is, the projected data $\sum_{i=1}^{D} g_i z_i$ also follow an $\alpha$-stable distribution with the scale parameter being the $l_\alpha$ norm of the vector $[g_1, g_2, ..., g_D]^{\mathrm{T}}$.

Therefore, given two vectors $u_1, u_2 \in \mathbb{R}^D$ (e.g., $u_1$ and $u_2$ are the leading two rows in the data matrix $\mathbf{A}$), if $v_1 = \mathbf{R}^{\mathrm{T}} u_1$ and $v_2 = \mathbf{R}^{\mathrm{T}} u_2$, where the entries of $\mathbf{R} \in \mathbb{R}^{D \times k}$, $r_{ij}$, are i.i.d. samples of $S(\alpha, 1)$, then $x_j = v_{1,j} - v_{2,j}$, $j = 1, 2, ..., k$ are i.i.d. $S(\alpha, d_{(\alpha)})$, where $d_{(\alpha)}$ is the $l_\alpha$ distance between $u_1$ and $u_2$.

Thus, the problem eventually boils down to estimating the scale parameter of $S(\alpha, d_{(\alpha)})$, from $k$ i.i.d. samples. Estimators based on the maximum likelihood, which are asymptotically (as $k \to \infty$) optimal, are computationally very intensive except for $\alpha = 2, 1, 0+$; and hence they are not practical for many applications. We recommend the estimators based on the geometric mean which are computationally convenient and still quite accurate, especially when $\alpha$ is around 1. Moreover, the geometric mean estimators are convenient for theoretical analysis, e.g., variances and tail bounds.

Our main contributions include (A): Very sparse stable random projections; (B): Estimators and tail bounds for stable random projections.

### 2.1 Very Sparse Stable Random Projections

We suggest replacing the entries, $S(\alpha, 1)$, in the projection matrix $\mathbf{R}$ with the following i.i.d. entries

$$r_{ij} = \begin{cases} P_\alpha & \text{with prob. } \frac{\beta}{2} \\ 0 & \text{with prob. } 1 - \beta \\ -P_\alpha & \text{with prob. } \frac{\beta}{2} \end{cases}, \quad (4)$$

where $P_\alpha$ denotes an $\alpha$-Pareto variable, $Pareto(\alpha, 1)$, $\mathbf{Pr}(P_\alpha > t) = \frac{1}{t^\alpha}$ if $t \geq 1$; and 0 otherwise. The projected data $\sum_{i=1}^{D} g_i r_{ij}$ will be *asymptotically stable* under certain regularity condition. This procedure is beneficial because

- It is much easier to sample from an $\alpha$-Pareto distribution than from $S(\alpha, 1)$.
- Computing $\mathbf{A} \times \mathbf{R}$ costs only $O(\beta nDk)$ as opposed to $O(nDk)$, a $\frac{1}{\beta}$-fold speedup, where the data matrix $\mathbf{A} \in \mathbb{R}^{n \times D}$.
- The storage (of $\mathbf{R}$) cost is reduced from $O(Dk)$ to $O(\beta Dk)$.

We will give the conditions for convergence and the rates of convergence as functions of $\alpha$, $\beta$, and the data regularity conditions. Two "easy-to-remember" statements are:

- In order for very sparse stable random projections to converge, the data should have at least bounded $\alpha$th moments.
- When the data have bounded second moments and $\alpha \leq 1$, we can let $\beta = \frac{1}{\sqrt{D}}$ and the rate of convergence will be at least $O\left(D^{-1/2}\right)$, which is fast even for moderate $D$.

We notice that [39, 19] had suggested using (4) with $\beta = 1$ as the standard practice, without showing the convergence conditions and the rates of convergence.

Non-asymptotic analysis on very sparse stable random projections is difficult even for $\beta = 1$. Therefore, whenever we discuss about estimators and tail bounds, we assume that we are using regular stable random projections.

## 2.2 Estimators and Tail Bounds

Here we only present the estimator for $d_{(\alpha)}$, the $l_\alpha$ distance between $u_1$ and $u_2 \in \mathbb{R}^D$, from the projected data difference, $x_j = (v_{1,j} - v_{2,j}) \sim S(\alpha, d_{(\alpha)}), j = 1, 2, ..., k$. Our proposed estimator is based on the geometric mean:

$$\hat{d}_{(\alpha),gm} = \frac{\prod_{j=1}^k |x_j|^{1/k}}{\left[\frac{2}{\pi}\Gamma\left(\frac{1}{k}\right)\Gamma\left(1 - \frac{1}{k\alpha}\right)\sin\left(\frac{\pi}{2}\frac{1}{k}\right)\right]^k}, \quad x_j \sim S(\alpha, d_{(\alpha)}), \text{ i.i.d.}, \quad k > \frac{1}{\alpha} \quad (5)$$

- $\hat{d}_{(\alpha),gm}$ is unbiased, i.e., $\mathrm{E}\left(\hat{d}_{(\alpha),gm}\right) = d_{(\alpha)}$.
- The correction term can be pre-computed for small $k$. For large $k$, we have the asymptotic formula

$$\left[\frac{2}{\pi}\Gamma\left(\frac{1}{k}\right)\Gamma\left(1 - \frac{1}{k\alpha}\right)\sin\left(\frac{\pi}{2}\frac{1}{k}\right)\right]^k \to \exp\left(-\gamma_e\left(1 - \frac{1}{\alpha}\right)\right), \quad (6)$$

where $\gamma_e = 0.577215665...$, is the Euler's constant. It converges from above monotonically.

- The variance is (valid for $k > \frac{2}{\alpha}$)

$$\mathrm{Var}\left(\hat{d}_{(\alpha),gm}\right) = d_{(\alpha)}^2 \left\{ \frac{\left[\frac{2}{\pi}\Gamma\left(\frac{2}{k}\right)\Gamma\left(1 - \frac{2}{k\alpha}\right)\sin\left(\pi\frac{1}{k}\right)\right]^k}{\left[\frac{2}{\pi}\Gamma\left(\frac{1}{k}\right)\Gamma\left(1 - \frac{1}{k\alpha}\right)\sin\left(\frac{\pi}{2}\frac{1}{k}\right)\right]^{2k}} - 1 \right\} = d_{(\alpha)}^2 \left(\frac{\pi^2}{6k}\left(\frac{1}{2} + \frac{1}{\alpha^2}\right) + O\left(\frac{1}{k^2}\right)\right). \quad (7)$$

- The tail bounds are

$$\mathbf{Pr}\left(\hat{d}_{(\alpha),gm} - d_{(\alpha)} > \epsilon d_{(\alpha)}\right) \leq \exp\left(-k\frac{\epsilon^2}{M_{R,\alpha,\epsilon}}\right), \quad \epsilon > 0, \quad k > \frac{1}{\alpha}, \quad (8)$$

$$\mathbf{Pr}\left(\hat{d}_{(\alpha),gm} - d_{(\alpha)} < -\epsilon d_{(\alpha)}\right) \leq \exp\left(-k\frac{\epsilon^2}{M_{L,\alpha,\epsilon,k_0}}\right), \quad 0 < \epsilon \leq 1, k > k_0 > \frac{1}{\alpha}, \quad (9)$$

where $M_{R,\alpha,\epsilon}$ and $M_{L,\alpha,\epsilon,k_0}$ are explicitly given.

- Using $\hat{d}_{(\alpha),gm}$, it suffices to let $k = O\left(\frac{\log(n)}{\epsilon^2}\right)$ so that the $l_\alpha$ distance between any pair of data points (or data streams) among $n$ data points (or data streams) can be estimated within a $1 \pm \epsilon$ factor, with high probability. The constant can be determined from $M_{R,\alpha,\epsilon}$ and $M_{L,\alpha,\epsilon,k_0}$.

## 3 Very Sparse Stable Random Projections

We suggest a procedure to simplify stable random projections and significantly reduce the processing and storage cost.

Recall the basic fact about stable distributions: If $z_1, z_2, ..., z_D$, i.i.d. are $S(\alpha, 1)$, then for any constants (i.e., the original data) $g_1, g_2, ..., g_D$, we have $\sum_{i=1}^D g_i z_i \sim S\left(\alpha, (\sum_{i=1}^D |g_i|^\alpha)^{1/\alpha}\right)$. That is, the projected data $\sum_{i=1}^D g_i z_i$ also follow an $\alpha$-stable distribution with the scale parameter being the $l_\alpha$ norm of the vector $[g_1, g_2, ..., g_D]^\mathrm{T}$.

However, it is expensive to sample from $S(\alpha, 1)$, if $\alpha \neq 1$ or 2. For example, [55, Proposition 1.71.1] describes a popular procedure for sampling from $S(\alpha, 1)$. That is, we first sample $W_1$ uniform on $(-\frac{\pi}{2}, \frac{\pi}{2})$ and $E_1$ from an exponential distribution with mean 1. If $W_1$ and $E_1$ are independent, then

$$\frac{\sin(\alpha W_1)}{\cos(W_1)^{1/\alpha}}\left(\frac{\cos\left((1-\alpha)W_1\right)}{E_1}\right)^{(1-\alpha)/\alpha} \quad (10)$$

is distributed as $S(\alpha, 1)$. Apparently, this procedure is quite costly.

The procedure for conducting stable random projections is also quite expensive. For example, the cost of matrix multiplication $\mathbf{A} \times \mathbf{R}$ would be $O(nDk)$, where $\mathbf{A} \in \mathbb{R}^{n \times D}$ is the data matrix and $\mathbf{R} \in \mathbb{R}^{D \times k}$ is the projection matrix consisting of i.i.d. samples of $S(\alpha, 1)$.

There is also a considerable storage cost for $\mathbf{R}$. There are at least two reasons why we need to store $\mathbf{R}$. Firstly, in some scenarios, we need to consider that new data points (or data streams) will be added to the dataset. Secondly, in data stream computations, the data entries do not necessarily arrive in orders[39]. In fact, the data may be also subject to frequent additions/subtractions[18, 19]. The cost of storing $\mathbf{R}$ is $O(Dk)$.

To tackle the above issues, we suggest replacing $S(\alpha, 1)$ with the following

$$z_i = \begin{cases} P_\alpha & \text{with prob. } \frac{\beta}{2} \\ 0 & \text{with prob. } 1 - \beta \\ -P_\alpha & \text{with prob. } \frac{\beta}{2} \end{cases}, \tag{11}$$

where $P_\alpha$ denotes an $\alpha$-Pareto distribution, $Pareto(\alpha, 1)$. That is, $\mathbf{Pr}(P_\alpha > t) = \frac{1}{t^\alpha}$ if $t \geq 1$; and 0 otherwise.

We call this approach *very sparse stable random projections* because on average only $\beta$ fraction of the entries are non-zeros, i.e., a $\frac{1}{\beta}$-fold speedup in computing $\mathbf{A} \times \mathbf{R}$, from $O(nDk)$ down to $O(\beta nDk)$. The storage cost is reduced from $O(Dk)$ to $O(\beta Dk)$.

There are two fundamental reasons why this approach should work:

- The data should satisfy certain *regularity* conditions otherwise the $l_\alpha$ norms may not be meaningful. For example, when using the $l_1$ norm, implicitly we expect that the data have at least bounded first moments.
- The data dimension $D$ should be very large, otherwise there would be no need for approximate answers.

We are inspired by the recent work on *very sparse random projections* for dimension reduction in $l_2$ [48], which showed the regularity condition for convergence and rate of convergence using known statistical theorems: the Lindeberg central limit theorem and the Berry-Esseen theorem. In our case, we also need to analyze under what conditions *very sparse stable random projections* will converge, as well as the rates of convergence. The necessary and sufficient condition for convergence is known (e.g., [36]), for both i.i.d. and non-i.i.d. (independent but not identical) cases. The rates of convergence for the i.i.d. case are also known, see [33] and a recent paper[44]. In our case, since we have to deal with the non-i.i.d. scenario, we will resort to the first principle, i.e., by studying the characteristic function of $\sum_{i=1}^{D} g_i z_i$, in Lemma 1 (proved in Appendix A).

**Lemma 1** *Suppose $z_i$, $i = 1, 2, ..., D$, are i.i.d. random variables defined in (11). Then as $D \to \infty$,*

$$\frac{\sum_{i=1}^{D} z_i g_i}{\left(\beta \sum_{s=1}^{D} |g_s|^\alpha\right)^{\frac{1}{\alpha}}} \to S\left(\alpha, \left(\Gamma(1-\alpha)\cos\left(\frac{\pi}{2}\alpha\right)\right)^{1/\alpha}\right), \quad \text{in distribution,} \tag{12}$$

*provided*

$$\frac{\max_{1 \leq i \leq D}(|g_i|)}{\left(\sum_{s=1}^{D} |g_s|^\alpha\right)^{\frac{1}{\alpha}}} \to 0. \tag{13}$$

*The rate of convergence is*

$$\begin{cases} O\left(\frac{\sum_{i=1}^{D}|g_i|^2}{\beta^{2/\alpha-1}\left(\sum_{i=1}^{D}|g_i|^\alpha\right)^{2/\alpha}}\right), & \text{if } 1 \leq \alpha < 2 \\ O\left(\max\left\{\frac{\sum_{i=1}^{D}|g_i|^{2\alpha}}{\left(\sum_{i=1}^{D}|g_i|^\alpha\right)^2}, \frac{\sum_{i=1}^{D}|g_i|^2}{\beta^{2/\alpha-1}\left(\sum_{i=1}^{D}|g_i|^\alpha\right)^{2/\alpha}}\right\}\right), & \text{if } 0 < \alpha < 1 \end{cases}. \tag{14}$$

There is no need to consider $\alpha = 2$ because we can sample from normals or the sparse distribution suggested in [1, 48]. Note that (13) is only a convenient sufficient condition.

Lemma 1 is not very interpretable. For convenience, we will assume that the data $g_i$'s are i.i.d. Suppose the data have bounded second moments, we have the following corollary (can be shown by strong law of large numbers).

**Corollary 1** *Suppose $|g_i|$'s are i.i.d. with bounded second moments, then the convergence condition (13) is satisfied. If we choose $\beta = \frac{1}{\sqrt{D}}$, then the rate of convergence is $O\left(D^{-\min(1, 1/\alpha - 1/2)}\right)$.*

In other words, if the data have bounded second moments (not a very strict condition), we can achieve a significant $\sqrt{D}$-fold speedup and the rate of convergence is still reasonable if $\alpha$ is not close to 2. For example, the rate is

$O\left(D^{-1/2}\right)$ when $\alpha = 1$. In practical applications, because $D$ is very large, a rate $O\left(D^{-1/2}\right)$ should be fast enough. On the other hand, when $\alpha$ is approaching 2, then the rate of convergence will be very slow (if converges at all) even if we let $\beta = 1$. Therefore, we do not recommend replacing the stable distribution with Pareto when $\alpha$ is close to 2.

Next, we will consider the case when the data do not have bounded second moments or even first moments. To simplify the arguments, we assume the data $|g_i|$'s are i.i.d. and follow an $\eta$-Pareto distribution with $\eta < 2$. Recall if a random variable $x$ follows an $\eta$-Pareto distribution, then $\mathrm{E}(x^\gamma) < \infty$ if $\gamma < \eta$ and $\mathrm{E}(x^\gamma) = \infty$ if $\gamma \geq \eta$.

Heavy-tailed data (without bounded second moments) are usually modeled by Pareto-type distributions. [53] measured the $\eta$ values for many kinds of datasets. While it is quite often that $1 < \eta < 2$, it is not very common that $\eta < 1$. For example, [53] measured the word frequency has $\eta = 1.2$, which is the well-known highly heavy-tailed case. The frequency of family names has $\eta = 0.94$ and the intensity of wars has $\eta = 0.8$, both not too far from 1.

**Corollary 2** *Suppose $|g_i|$'s are i.i.d. $\eta$-Pareto with $\eta < 2$, then the convergence condition (13) is satisfied if $\eta > \alpha$. Assuming $\eta > \alpha$, the rate of convergence would be*

$$\begin{cases} O\left(\frac{1}{\beta^{2/\alpha-1}D^{2/\alpha-2/\eta}}\right), & \text{if} \quad 1 \leq \alpha < 2 \\ O\left(\max\left\{\frac{1}{D^{2-\max(1,2\alpha/\eta)}}, \frac{1}{\beta^{2/\alpha-1}D^{2/\alpha-2/\eta}}\right\}\right), & \text{if} \quad 0 < \alpha < 1 \end{cases} \quad (15)$$

*If we choose $\beta = D^{-\frac{1-\alpha/\eta}{2-\alpha}}$, then the rate of convergence would be*

$$\begin{cases} O\left(\frac{1}{D^{1/\alpha-1/\eta}}\right), & \text{if} \quad 1 \leq \alpha < 2 \\ O\left(\max\left\{\frac{1}{D^{2-\max(1,2\alpha/\eta)}}, \frac{1}{D^{1/\alpha-1/\eta}}\right\}\right), & \text{if} \quad 0 < \alpha < 1 \end{cases} \quad (16)$$

**Proof:** *This corollary can be shown by the fact that if $x_i$ is $\eta$-Pareto, i.i.d., then $\sum_{i=1}^{D} x_i^\gamma$ grows as $O\left(D^{\max(1,\gamma/\eta)}\right)$. See [28, Example 2.7.4].*

Therefore, in order for very sparse stable random projections to converge (for any $0 < \beta \leq 1$), we have to make sure that the original data should have at least bounded $\alpha$th moment. This is a very natural requirement. When the data have bounded higher moments, we can obtain a faster rate of convergence and afford a smaller $\beta$.

We can see that if $D$ is larger enough (say $10^5$), it is quite easy to achieve a 10-fold or 100-fold speedup. A factor of 100 (or even 10) may be significant enough to make a theoretically appealing algorithm become a practical one.

Our numerical studies show that very sparse stable random projections work really well (probably more than what we would expect). In the next three subsections, we will present some numerical results on the synthetic data, some Web crawl data, and the Harvard microarray data, respectively, for the $l_1$ case ($\alpha = 1$).

### 3.1 Numerical Results on Synthetic Data

We simulate data from $Pareto(\eta, 1)$ for $\eta = 1.5$ and $\eta = 2.0$. We choose $\alpha = 1.0$ (i.e., the $l_1$ norm). Corollary 2 recommends $\beta = D^{-\frac{1-\alpha/\eta}{2-\alpha}}$, which is respectively $D^{-1/3}$ and $D^{-1/2}$ for $\eta = 1.5$ and $\eta = 2.0$. To make the results more interesting, we choose $\beta = D^{-0.4}$ and $D^{-0.75}$, respectively, otherwise all curves will simply overlap.

We generate data for $D$ ranging from 100 to $10^6$ and apply very sparse stable random projections ($\beta = D^{-0.4}$ and $D^{-0.75}$) for $k$ ranging from 10 to 100. We then estimate $l_1$ norms using the geometric mean estimator we will discuss in Section 4, as if the projected data were exactly stable. The mean square errors (MSE's) are presented in Figure 1, which only plots $D = 100, 500,$ and $1000$ because the curves corresponding to larger $D$'s overlap. The results indicate that very sparse stable random projections work really well even when $D$ is not too large.

### 3.2 Numerical Results on Web Crawl Data

We apply very sparse stable random projections on some MSN Web crawl data. We pick two pairs of words, THIS-HAVE, and SCHOOL-PROGRAM. The data dimension $D = 2^{16} = 65536$. For each word, the $i$th entry ($i = 1$ to $D$) is the number of occurrences this word appeared in the $i$th Web page. It is well-known that the word frequency data are highly heavy-tailed. The pair THIS-HAVE are frequent words while the pair SCHOOL-PROGRAM are relatively infrequent. Some summary statistics are given in Table 1, which verifies that the data are indeed highly heavy-tailed, especially the pair SCHOOL-PROGRAM.

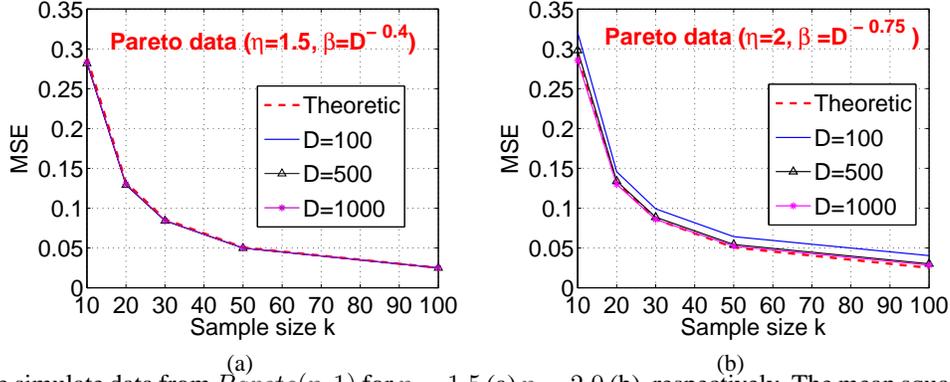

Figure 1: We simulate data from $\tilde{P}areto(\eta, 1)$ for $\eta = 1.5$ (a) $\eta = 2.0$ (b), respectively. The mean square errors (MSE) are plotted against the sample size $k$ for each $D$=100, 500, and 1000. The "theoretic" curves were the theoretical variances assuming the data are exactly (instead of asymptotically) stable (see Lemma 2).

Table 1: For each pair of words ($u_1$ v.s. $u_2$), we compute the difference vector $u = u_1 - u_2$ and ratios of the empirical moments for illustrating that the data are highly heavy-tailed.

|  | Sparsity ($u_1$) | Sparsity ($u_2$) | Sparsity ($u$) | $\frac{E(|u|^2)}{(E(|u|))^2}$ | $\frac{E(|u|^4)}{(E(|u|^2))^2}$ |
|---|---|---|---|---|---|
| THIS ($u_1$) - HAVE ($u_2$) | 0.4226 | 0.2674 | 0.4378 | 9.97 | 239.81 |
| SCHOOL ($u_1$) - PROGRAM ($u_2$) | 0.0695 | 0.0816 | 0.1279 | 51.77 | 4076.30 |

For each pair, we estimate the $l_1$ distance using very sparse stable random projections with $\beta = 0.1, 0.01,$ and $0.001$. The results are presented in Figure 2. For the pair THIS-HAVE, even when $\beta = 0.001$, the results are indistinguishable from what we would obtain by exact stable random projections. For the pair SCHOOL-PROGRAM, when $\beta = 0.01$, the results are good. However, when $\beta = 0.001$, we see considerably larger errors. This is because the data are sparse (sparsity = 0.1279, meaning that the "effective data dimension" should be much smaller than $D = 65536$) and the data are highly heavy-tailed. Note that $\frac{1}{\sqrt{D}} = 0.039$, $\frac{1}{D^{0.4}} = 0.0118$.

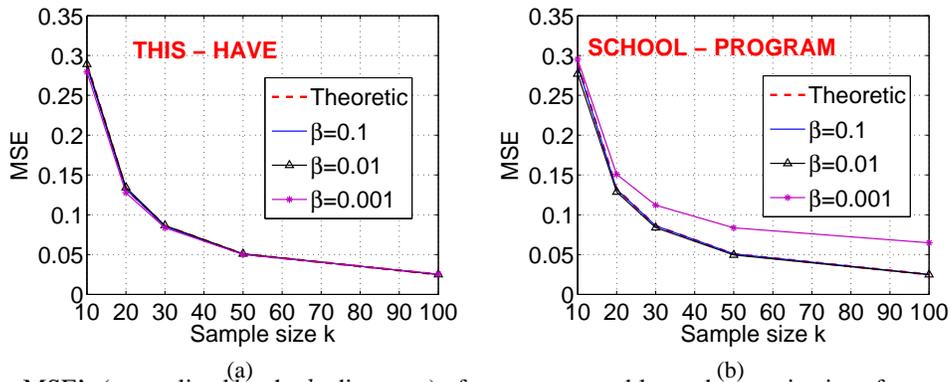

Figure 2: The MSE's (normalized by the $l_1$ distances) of very sparse stable random projections for two pairs of words. The "theoretic" curves were the theoretical variances assuming the data are exactly stable (see Lemma 2).

## 3.3 Numerical Results on Classifying Microarray Data

Usually the purpose of computing distances is for the subsequent tasks such as clustering, classification, information retrieval, etc. Here we consider the task of classifying deceases in the Harvard microarray dataset[9][1]. The original dataset contains 203 samples (specimens) in 12600 gene dimensions, including 139 lung adenocarcinomas (12 of which may be suspicious), 17 normal samples, 20 pulmonary carcinoids, 21 sqamous cell lung carcinomas, and 6 SCLC cases. We select the first three classes (in total 139 + 17+20 = 176 samples) as our test dataset. For each specimen, we subtract the median (across genes). However, we did not perform any normalization.

A simple nearest neighbor classifier can classify the samples almost perfectly using the $l_1$ distance. When $m = 1, 3, 5, 7, 9$, the $m$-nearest neighbor classifier mis-classifies 3, 2, 2, 2, 2, samples, respectively. For comparisons, using the ($l_2$) correlation distance (i.e., 1 - correlation coefficient), when $m = 1, 3, 5, 7, 9$, the $m$-nearest neighbor classifier mis-classifies 6, 4, 4, 4, 5, samples, respectively.

We conduct both stable (i.e., Cauchy) random projections and very sparse stable random projections ($\beta = 0.1$, 0.01, and 0.001) on the dataset and classify the specimens using a 5-nearest neighbor classifier based on the $l_1$ distances. Figure 3 indicates that (A): stable random projections can achieve similar classification accuracy using about 100 projections (as opposed to the original $D = 12600$ dimensions); (B): very sparse stable random projections work well when $\beta = 0.1$ and 0.01. Even with $\beta = 0.001$, the classification results are only slightly worse.

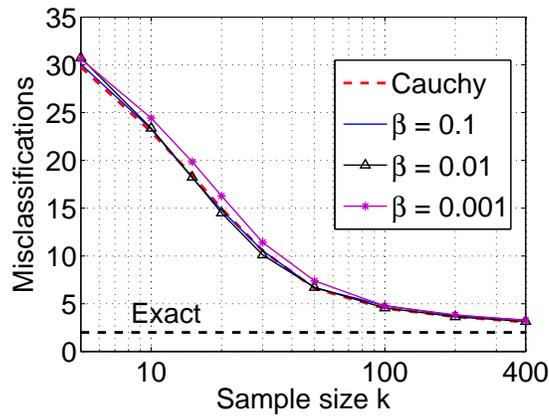

Figure 3: We apply stable (Cauchy) random projections and very sparse stable random projections ($\beta = 0.1$, 0.01, and 0.001) on the Harvard microarray dataset ($n = 176$ specimens in $D = 12600$ dimensions) and classify the specimens using a 5-nearest neighbor classifier based on the $l_1$ distances. The horizontal dashed line indicates that, using the exact $l_1$ distances, a 5-nearest neighbor classifier mis-classifies 2 samples. The other curves show that using stable random projections we can achieve almost the same misclassification errors with just 100 projections (as opposed to the original 12600 dimensions). Very sparse stable random projections with $\beta = 0.1$ and 0.01 perform almost indistinguishably from Cauchy random projections. Even when $\beta = 0.001$, the results are only slightly worse. Each curve is averaged over 100 runs.

## 3.4 A More General Scheme for Sparse Stable Random Projections

Instead of using the sparse distribution in (11), we could, alternatively, consider the following more general form:

$$z_i = \begin{cases} P_{\alpha,\mu} & \text{with prob. } \frac{\beta}{2} \\ 0 & \text{with prob. } 1 - \beta \\ -P_{\alpha,\mu} & \text{with prob. } \frac{\beta}{2} \end{cases}, \qquad (17)$$

where $P_{\alpha,\mu}$ denotes $Pareto(\alpha, \mu)$. That is, $\mathbf{Pr}\left(P_{\alpha,\mu} > t\right) = \frac{\mu^\alpha}{t^\alpha}$ if $t \geq \mu$; and 0 otherwise.

Theoretically, (17) is appealing. When we choose a smaller (than 1) $\mu$, e.g., $\mu = 1/D^\gamma$ for some $\gamma > 0$, we can actually achieve a faster rate of convergence and a less restrictive condition for ensuring convergence.

However, as $\mu$ decreases, the probability density function (PDF) of $P_{\alpha,\mu}$ becomes more steep near $\mu$ (as shown in Figure 4), i.e., harder to obtain random samples of good quality.

---
[1] http://research.dfci.harvard.edu/meyersonlab/lungca/files/DatasetA_12600gene.txt

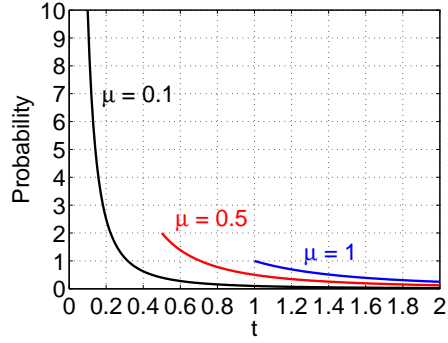

Figure 4: The probability density functions of $P_{\alpha,\mu}$ for $\alpha = 1$ and $\mu = 1, 0.5$, and $0.1$.

To conclude this section, we should point out that non-asymptotic analysis on very sparse stable random projections is difficult. For the rest of the paper, we will assume that we use the regular stable random projections.

## 4 The Geometric Mean Estimators and Tail Bounds

We present the results for estimating $d_{(\alpha)}$ and $d_{(\alpha)}^\alpha$, where $d_{(\alpha)}$ is the $l_\alpha$ distance between $u_1$ and $u_2 \in \mathbb{R}^D$, from the projected data $v_1$ and $v_2$. $v_1 = \mathbf{R}^\mathrm{T} u_1$ and $v_2 = \mathbf{R}^\mathrm{T} u_2$, where $\mathbf{R} \in \mathbb{R}^{D \times k}$ have i.i.d. entries $r_{ij} \sim S(\alpha, 1)$. We always denote $x_j = v_{1,j} - v_{2,j}$, $j = 1, 2, ..., k$, $x_j \sim S(\alpha, d_{(\alpha)})$.

We develop two sets of estimators and tail bounds, one for estimating $d_{(\alpha)}$ and another for estimating $d_{(\alpha)}^\alpha$.

### 4.1 The Geometric Mean Estimator and Tail Bounds for $d_{(\alpha)}$

Our proposed geometric mean estimator $\hat{d}_{(\alpha),gm}$ for $d_{(\alpha)}$ is based on the following fact about $S(\alpha, d_{(\alpha)})$.

**Proposition 1** *Suppose $x \sim S(\alpha, d_{(\alpha)})$. Then for $-1 < \lambda < \alpha$,*

$$E\left(|x|^\lambda\right) = d_{(\alpha)}^\lambda \frac{2}{\pi} \Gamma\left(1 - \frac{\lambda}{\alpha}\right) \Gamma(\lambda) \sin\left(\frac{\pi}{2}\lambda\right). \tag{18}$$

***Proof*** *Although not explicitly stated in [60], one can infer this result by combining various statements in [60] (page 63, page 116, 117. Note the typos in (2.1.9) and (2.1.10)).*

*For the sake of verification, we derive $E\left(|x|^\lambda\right)$ for $0 \leq \lambda < \alpha$, by completing the result stated in [55, Property 1.2.17, page 18], which says*

$$E\left(|x|^\lambda\right) = d_{(\alpha)}^\lambda \frac{2^{\lambda-1} \Gamma\left(1 - \frac{\lambda}{\alpha}\right)}{\lambda \int_0^\infty \frac{\sin^2 u}{u^{\lambda+1}} du}, \qquad 0 \leq \lambda < \alpha. \tag{19}$$

*We can find the explicit expression for the integral[37, 3.823, page 484], as*

$$\int_0^\infty \frac{\sin^2 u}{u^{\lambda+1}} du = -\frac{\Gamma(-\lambda) \cos\left(\frac{\pi}{2}\lambda\right)}{2^{-\lambda+1}} \tag{20}$$

*By Euler's reflection formula, $\Gamma(1-z)\Gamma(z) = \frac{\pi}{\sin(\pi z)}$, we know $\Gamma(-\lambda) = -\frac{\pi}{\sin(\pi\lambda)} \frac{1}{\Gamma(\lambda+1)}$, and the desired result follows after some algebra.*

We need the above result for $\lambda < \alpha$ in order to derive the proposed unbiased estimator. We will need to consider $\lambda < 0$ for proving two-sided tail bounds.

From (18), we can design an unbiased estimator for $d_{(\alpha)}$ as (by taking $\lambda = 1/k$)

$$\hat{d}_{(\alpha),gm} = \frac{\prod_{j=1}^k |x_j|^{1/k}}{\left[\frac{2}{\pi} \Gamma\left(\frac{1}{k}\right) \Gamma\left(1 - \frac{1}{k\alpha}\right) \sin\left(\frac{\pi}{2}\frac{1}{k}\right)\right]^k}, \qquad x_j \sim S(\alpha, d_{(\alpha)}), \text{ i.i.d.}, \quad k > \frac{1}{\alpha}. \tag{21}$$

The denominator in $\hat{d}_{(\alpha),gm}$ can be pre-computed and it converges to a fixed value depending on $\alpha$. We call this estimator the *geometric mean estimator* although it is really the geometric mean with corrections for both $k$ and $\alpha$.

We prove some useful properties of $\hat{d}_{(\alpha),gm}$ in the following Lemma (see the proof in Appendix B)

**Lemma 2** *The estimator, $\hat{d}_{(\alpha),gm}$, as defined in (21)*

- *It is unbiased, i.e., $E\left(\hat{d}_{(\alpha),gm}\right) = d_{(\alpha)}$.*

- *As $k \to \infty$,*

$$\left[\frac{2}{\pi}\Gamma\left(\frac{1}{k}\right)\Gamma\left(1-\frac{1}{k\alpha}\right)\sin\left(\frac{\pi}{2}\frac{1}{k}\right)\right]^k \to \exp\left(-\gamma_e\left(1-\frac{1}{\alpha}\right)\right), \tag{22}$$

*where $\gamma_e = 0.577215665...$, is the Euler's constant. It converges from above monotonically.*

- *The variance of $\hat{d}_{(\alpha),gm}$ is, provided $k > \frac{2}{\alpha}$,*

$$\operatorname{Var}\left(\hat{d}_{(\alpha),gm}\right) = d_{(\alpha)}^2 \left\{ \frac{\left[\frac{2}{\pi}\Gamma\left(\frac{2}{k}\right)\Gamma\left(1-\frac{2}{k\alpha}\right)\sin\left(\pi\frac{1}{k}\right)\right]^k}{\left[\frac{2}{\pi}\Gamma\left(\frac{1}{k}\right)\Gamma\left(1-\frac{1}{k\alpha}\right)\sin\left(\frac{\pi}{2}\frac{1}{k}\right)\right]^{2k}} - 1 \right\} = d_{(\alpha)}^2 \left(\frac{\pi^2}{6k}\left(\frac{1}{2}+\frac{1}{\alpha^2}\right) + O\left(\frac{1}{k^2}\right)\right). \tag{23}$$

In Lemma 2, the asymptotic properties of $\hat{d}_{(\alpha),gm}$ are convenient (and more interpretable) when we derive the tail bounds for $\hat{d}_{(\alpha),gm}$. The monotonicity property will also be used a couple of times. These interesting asymptotic properties are proved by careful Taylor expansions and using the infinite-product representations of the Gamma function and the sine function.

In Appendix C, we prove Lemma 3 for the tail bounds of $\hat{d}_{(\alpha),gm}$.

**Lemma 3** *The right tail bound*

$$\mathbf{Pr}\left(\hat{d}_{(\alpha),gm} - d_{(\alpha)} > \epsilon d_{(\alpha)}\right) \le \exp\left(-k\frac{\epsilon^2}{M_{R,\alpha,\epsilon}}\right), \quad \epsilon > 0, \quad k > \frac{1}{\alpha}, \tag{24}$$

*where*

$$\frac{1}{M_{R,\alpha,\epsilon}} = \frac{1}{c_\alpha \epsilon}\log(1+\epsilon) - \frac{1}{\epsilon^2}\log\left(\frac{2}{\pi}\Gamma\left(1-\frac{\epsilon}{c_\alpha\alpha}\right)\Gamma\left(\frac{\epsilon}{c_\alpha}\right)\sin\left(\frac{\pi}{2}\frac{\epsilon}{c_\alpha}\right)\right) - \frac{1}{c_\alpha\epsilon}\gamma_e\left(1-\frac{1}{\alpha}\right) \tag{25}$$

$$c_\alpha = \frac{\pi^2}{6}\left(\frac{1}{2}+\frac{1}{\alpha^2}\right) \tag{26}$$

*The left tail bound*

$$\mathbf{Pr}\left(\hat{d}_{(\alpha),gm} - d_{(\alpha)} < -\epsilon d_{(\alpha)}\right) \le \exp\left(-k\frac{\epsilon^2}{M_{L,\alpha,\epsilon,k_0}}\right), \quad 0 < \epsilon \le 1, k > k_0 > \frac{1}{\alpha}, \tag{27}$$

*where*

$$\frac{1}{M_{L,\alpha,\epsilon,k_0}} = -\frac{1}{c_\alpha\epsilon}\log(1-\epsilon) - \frac{1}{\epsilon^2}\log\left(-\frac{2}{\pi}\Gamma\left(1+\frac{\epsilon}{c_\alpha\alpha}\right)\Gamma\left(-\frac{\epsilon}{c_\alpha}\right)\sin\left(\frac{\pi}{2}\frac{\epsilon}{c_\alpha}\right)\right)$$
$$- \frac{1}{c_\alpha\epsilon}k_0\log\left(\left[\frac{2}{\pi}\Gamma\left(\frac{1}{k_0}\right)\Gamma\left(1-\frac{1}{k_0\alpha}\right)\sin\left(\frac{\pi}{2}\frac{1}{k_0}\right)\right]\right). \tag{28}$$

Note that in (28), $k_0 \log\left(\left[\frac{2}{\pi}\Gamma\left(\frac{1}{k_0}\right)\Gamma\left(1-\frac{1}{k_0\alpha}\right)\sin\left(\frac{\pi}{2}\frac{1}{k_0}\right)\right]\right)$ converges to $-\gamma_e\left(1-\frac{1}{\alpha}\right)$; and hence it is well-behaved. Restricting $k > k_0$ in the left tail shall not raise a concern. Recall our goal is to show that $k = O\left(\frac{\log(n)}{\epsilon^2}\right)$ suffices, which is usually not too small. We could adjust $k_0$ to match $M_{R,\alpha,\epsilon}$ and $M_{R,\alpha,\epsilon,k_0}$ so that we can have a convenient "symmetric" bound $\mathbf{Pr}\left(|\hat{d}_{(\alpha),gm} - d_{(\alpha)}| > \epsilon d_{(\alpha)}\right)$. We showed this for $\alpha = 1$ in a technical report [47].

We consider the techniques (in Appendix C) for deriving the tail bounds in Lemma 3 are interesting:

- First we note that $\hat{d}_{(\alpha),gm}$, which only has moments up to $k$, does not have a moment generating function; and hence we can not use the popular Chernoff bound [17]. However, we can always use the Markov moment bound. [54] has shown that the moment bound is always sharper than the Chernoff bound for positive random variables, even when the Chernoff bound does exist.
- To get the optimal moment bound in our case is difficult (unless $\alpha = 1$). We resort to sub-optimal (but asymptotically optimal) bounds by realizing that $\hat{d}_{(\alpha),gm}$ can be treated as a gamma random variable when $k$ is large enough and $\epsilon$ is small. For a gamma, we know its optimal moment bound[54, Example 3.3]. The "gamma approximation" is due to the central limit theorem for large $k$. For positively-skewed variables, it is usually a good idea to use gamma rather than normal as long as both have the same asymptotic first two moments[49].

The right tail bound in Lemma 3 is only "pseudo-exponential," because the constant $M_{R,\alpha,\epsilon}$ depends on $\epsilon$. However, for a given $\epsilon$, no matter how large it is, we can always find the upper bound in an exponential form. From a practical point of view, what really matters is that the constants should be as small as possible. Indeed, as illustrated in Figures 5 and 6, $M_{R,\alpha,\epsilon}$ and $M_{L,\alpha,\epsilon,k_0}$ are reasonably small.

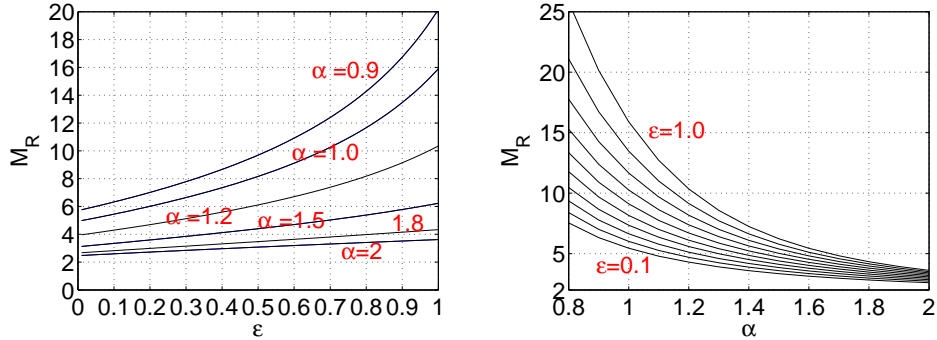

(a) (b)
Figure 5: We plot the right tail bound constant $M_{R,\alpha,\epsilon}$ in Lemma 3 for a certain range of $\alpha$ and $\epsilon$.

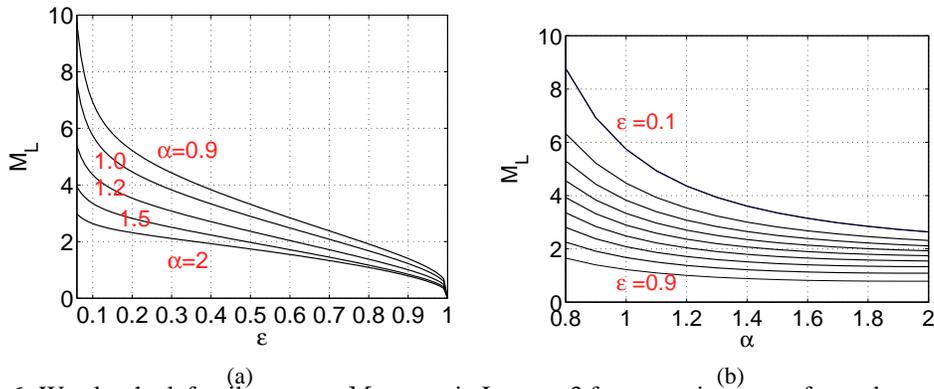

(a) (b)
Figure 6: We plot the left tail constant $M_{L,\alpha,\epsilon,k_0}$ in Lemma 3 for a certain range of $\alpha$ and $\epsilon$, at $k_0 = 100$.

A direct consequence of the tail bounds in Lemma 3 yields the following JL-like Lemma, which is weaker than the classical JL lemma, because the estimator $\hat{d}_{(\alpha),gm}$ is not a metric.

**Lemma 4** *Using the estimator $\hat{d}_{(\alpha),gm}$, for any fixed $\epsilon > 0$, we only need $k = O\left(\frac{\log(n)}{\epsilon^2}\right)$ to guarantee that the $l_\alpha$ distance between any pair of points among $n$ data points can be estimated within a factor of $1 \pm \epsilon$. Moreover, the constant can be explicitly characterized.*

### 4.2 The Geometric Mean Estimator and Tail Bounds for $d_{(\alpha)}^\alpha$

Sometimes we are more interested in $d_{(\alpha)}^\alpha$ than $d_{(\alpha)}$. For example, the classical JL Lemma for $l_2$ is presented in terms of the squared $l_2$ distance. [18, 19] approximated the Hamming norm using the $l_\alpha$ norm raised to $\alpha$th power, with very small $\alpha$. Obviously we could raise the estimator (e.g., $\hat{d}_{(\alpha),gm}$) to the $\alpha$th power, as treated in [18, 19]. We are concerned about the tail bounds, because a $(1 \pm \epsilon)^\alpha$ factor becomes a $1 \pm \epsilon^2$ factor when $\alpha = \epsilon$.

We can again design an unbiased estimator for $d_{(\alpha)}^\alpha$, discussed in the next lemma.

**Lemma 5** *The following estimator, denoted by $\hat{E}_{(\alpha)}$,*

$$\hat{E}_{(\alpha),gm} = \frac{\prod_{j=1}^k |x_j|^{\alpha/k}}{\left[\frac{2}{\pi}\Gamma\left(\frac{\alpha}{k}\right)\Gamma\left(1-\frac{1}{k}\right)\sin\left(\frac{\pi}{2}\frac{\alpha}{k}\right)\right]^k}, \quad k > 1, \tag{29}$$

*is unbiased, i.e., $E\left(\hat{E}_{(\alpha),gm}\right) = d_{(\alpha)}^\alpha$.*

*As $k \to \infty$,*

$$\left[\frac{2}{\pi}\Gamma\left(\frac{\alpha}{k}\right)\Gamma\left(1-\frac{1}{k}\right)\sin\left(\frac{\pi}{2}\frac{\alpha}{k}\right)\right]^k \to \exp\left(-\alpha\gamma_e\left(1-\frac{1}{\alpha}\right)\right), \tag{30}$$

*decreasing monotonically with increasing $k$.*

*The variance of $\hat{E}_{(\alpha)}$ would be, ($k > 2$),*

$$\text{Var}\left(\hat{E}_{(\alpha),gm}\right) = (d_{(\alpha)})^{2\alpha}\left\{\frac{\left[\frac{2}{\pi}\Gamma\left(\frac{2\alpha}{k}\right)\Gamma\left(1-\frac{2}{k}\right)\sin\left(\pi\frac{\alpha}{k}\right)\right]^k}{\left[\frac{2}{\pi}\Gamma\left(\frac{\alpha}{k}\right)\Gamma\left(1-\frac{1}{k}\right)\sin\left(\frac{\pi}{2}\frac{\alpha}{k}\right)\right]^{2k}} - 1\right\} = (d_{(\alpha)})^{2\alpha}\left(\alpha^2\frac{\pi^2}{6k}\left(\frac{1}{2}+\frac{1}{\alpha^2}\right) + O\left(\frac{1}{k^2}\right)\right). \tag{31}$$

*The right tail bound*

$$\mathbf{Pr}\left(\hat{E}_{(\alpha),gm} - d_{(\alpha)}^\alpha > \epsilon d_{(\alpha)}^\alpha\right) \leq \exp\left(-k\frac{\epsilon^2}{G_{R,\alpha,\epsilon}}\right), \quad \epsilon > 0, \quad k > 1, \tag{32}$$

*where*

$$\frac{1}{G_{R,\alpha,\epsilon}} = \frac{1}{\alpha^2 c_\alpha \epsilon}\log(1+\epsilon) - \frac{1}{\epsilon^2}\log\left(\frac{2}{\pi}\Gamma\left(1-\frac{\epsilon}{c_\alpha\alpha^2}\right)\Gamma\left(\frac{\epsilon}{c_\alpha\alpha}\right)\sin\left(\frac{\pi}{2}\frac{\epsilon}{c_\alpha\alpha}\right)\right) - \frac{1}{c_\alpha\alpha\epsilon}\gamma_e\left(1-\frac{1}{\alpha}\right) \tag{33}$$

$$c_\alpha = \frac{\pi^2}{6}\left(\frac{1}{2}+\frac{1}{\alpha^2}\right) \tag{34}$$

*The left tail bound*

$$\mathbf{Pr}\left(\hat{E}_{(\alpha),gm} - d_{(\alpha)}^\alpha < -\epsilon d_{(\alpha)}^\alpha\right) \leq \exp\left(-k\frac{\epsilon^2}{G_{L,\alpha,\epsilon,k_0}}\right), \quad 0 < \epsilon \leq 1, k > k_0 > 1, \tag{35}$$

*where*

$$\frac{1}{G_{L,\alpha,\epsilon,k_0}} = -\frac{1}{\alpha^2 c_\alpha \epsilon}\log(1-\epsilon) - \frac{1}{\epsilon^2}\log\left(-\frac{2}{\pi}\Gamma\left(1+\frac{\epsilon}{c_\alpha\alpha^2}\right)\Gamma\left(-\frac{\epsilon}{c_\alpha\alpha}\right)\sin\left(\frac{\pi}{2}\frac{\epsilon}{c_\alpha\alpha}\right)\right)$$
$$-\frac{1}{\alpha^2 c_\alpha \epsilon}k_0\log\left(\left[\frac{2}{\pi}\Gamma\left(\frac{\alpha}{k_0}\right)\Gamma\left(1-\frac{1}{k_0}\right)\sin\left(\frac{\pi}{2}\frac{\alpha}{k_0}\right)\right]\right). \tag{36}$$

We use the same technique for proving Lemma 5 as for proving Lemma 3; and hence we skip the proof.

As illustrated in Figure 7, For the reasonable range of $\epsilon$ (e.g., $\leq 0.5$), our tail bounds produce sensible results even when $\alpha$ is extremely small (e.g., 0.001).

### 4.3 Fisher Efficiency of the Geometric Mean Estimators

The geometric mean estimators are numerically straightforward, however, they are not optimal. The maximum likelihood estimators (MLE's) are asymptotically optimal, as the sample size increases to infinity. The Fisher efficiency of a proposed estimator is defined to be the asymptotic ratio of the variance of the corresponding MLE to the variance of the proposed estimator. For stable distributions, evaluating the MLE's are in general numerically very intensive because the probability density functions have to be numerically calculated, except for $\alpha = 2, 1, 0+$. We have mentioned previously that there is no need to use the geometric mean estimator when $\alpha = 2$. The technical report [47] discussed in detail the MLE when $\alpha = 1$. We will discuss the case $\alpha \to 0+$ in Section 5, which will report that the Fisher efficiency of the geometric mean estimator is $\frac{6}{\pi^2} = 0.6079$ as $\alpha \to 0+$.

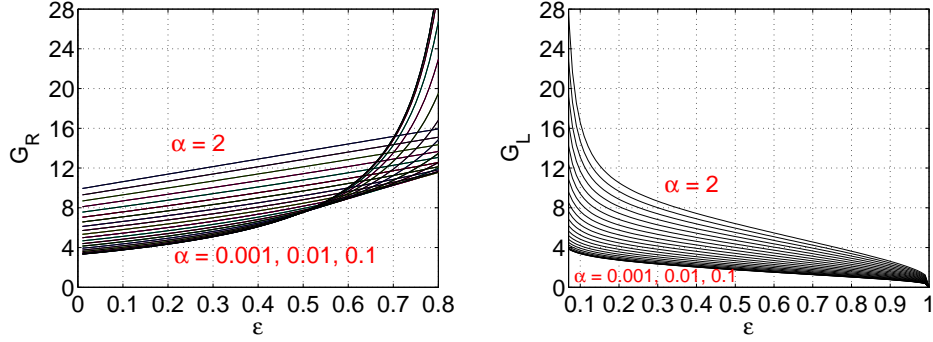

(a) $G_{R,\alpha,\epsilon}$ (b) $G_{L,\alpha,\epsilon,k_0}$

Figure 7: We plot the constants $G_{R,\alpha,\epsilon}$ and $G_{L,\alpha,\epsilon,k_0}$ ($k_0 = 100$) for the tail bounds in Lemma 5. The curves correspond to $\alpha = 0.001, 0.01$, and from 0.1 to 2 at an increment of 0.1.

[51] numerically evaluated the Fisher information (reciprocal of the asymptotic variance of the MLE) for a range of $\alpha$ values ($0.2 \leq \alpha \leq 2$) and commented on the numerical difficulty for smaller $\alpha$. Figure 8 shows that the geometric mean estimators are $\geq 80\%$ efficient when $\alpha = 0.4 \sim 1.1$. In other words, in the range of $\alpha = 0.4 \sim 1.1$, the geometric mean estimators are close to be optimal.

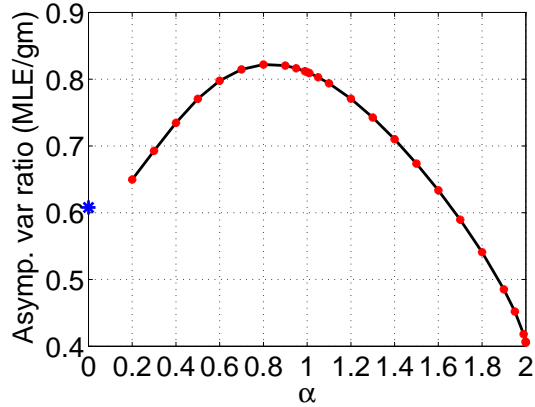

Figure 8: The Fisher efficiency of the geometric mean estimators based on the Fisher information reported in [51] and the asymptotic variances of the geometric mean estimators. The single point at (0,0.6079) is based our calculation in Section 5. Note that since the Fisher efficiency concerns only the asymptotic ratio of variances, the Fisher efficiency would be the same for both the geometric mean estimator for $d_{(\alpha)}$ and the geometric mean estimator for $d_{(\alpha)}^\alpha$.

## 5 Comparing Estimators (Especially when $\alpha \to 0+$)

The case $\alpha \to 0+$ is interesting and practically useful. For example, [18] and [20] applied stable random projections for approximating the Hamming norm and the max-dominance norm, respectively, using very small $\alpha$.

We will compare our proposed geometric mean estimator with the sample median estimator, which is a special case of sample quantile estimators[30, 31, 52], for the case when $\alpha \to 0+$. In general, the geometric mean estimator is considerably more accurate when the sample size $k$ is not very large. Asymptotically (as $k \to \infty$), the geometric mean estimator and sample median estimator is equivalent when $\alpha = 1$ (see the technical report[47]); but for any other specific $\alpha \neq 1$, we find the geometric mean estimator is always also more accurate asymptotically. At the end of this section, we will introduce a better estimator based on the maximum likelihood estimator, which has a simple form when $\alpha \to 0+$.

As shown by [21] (or see [46, 20]), if $x \sim S(\alpha, 1)$ and $\alpha \to 0+$, then $|x|^\alpha$ converges to $1/E_1$, where $E_1$ stands for an exponential distribution with mean 1. This result is quite obvious by taking limit of (10), the procedure for sampling from $S(\alpha, 1)$.

The median of $1/E_1$ is $1/\log(2)$. Therefore, given $x_j$, $j = 1, 2, ..., k$ i.i.d. $S(\alpha, d_{(\alpha)})$, when $\alpha \to 0+$, the sample median estimator of $d_{(\alpha)}^\alpha$ would be equivalent to

$$\hat{h}_{me} = \frac{\text{median}\{z_j, j = 1, 2, ..., k\}}{1/\log(2)}, \qquad (37)$$

where $z_j$'s are i.i.d. $\sim h/E_1$, $h = \lim_{\alpha \to 0+} d_{(\alpha)}^\alpha$, i.e., the hamming norm. In comparisons, the geometric mean estimator for $d_{(\alpha)}^\alpha$, derived in Section 4.2, becomes

$$\hat{h}_{gm} = \hat{E}_{(0+),gm} = \lim_{\alpha \to 0+} \frac{\prod_{j=1}^{k} |x_j|^{\alpha/k}}{\left[\frac{2}{\pi}\Gamma\left(\frac{\alpha}{k}\right)\Gamma\left(1 - \frac{1}{k}\right)\sin\left(\frac{\pi}{2}\frac{\alpha}{k}\right)\right]^k} = \frac{\prod_{j=1}^{k} z_j^{1/k}}{\Gamma\left(1 - \frac{1}{k}\right)^k}. \qquad (38)$$

We know the distribution of $z_j \sim h/E_1$ exactly:

$$\mathbf{Pr}(z_j \leq t) = \exp(-h/t), \qquad \mathbf{Pr}(z_j = t) = \exp(-h/t)\frac{h}{t^2}, \qquad t > 0. \qquad (39)$$

It is easy to show that for the geometric mean estimator,

$$\text{Var}\left(\hat{h}_{gm}\right) = h^2\left(\frac{\Gamma\left(1 - \frac{2}{k}\right)^k}{\Gamma\left(1 - \frac{1}{k}\right)^{2k}} - 1\right) = h^2\left(\frac{\pi^2}{6}\frac{1}{k}\right) + O\left(\frac{1}{k^2}\right). \qquad (40)$$

The asymptotic variance of the sample median estimator $\hat{h}_{me}$ can be shown using known statistical results on sample quantiles [56, Theorem 5.10].

$$\text{Var}\left(\hat{h}_{me}\right) = \frac{1}{4\mathbf{Pr}^2(z_j = h/\log(2))(1/\log(2))^2}\frac{1}{k} + O\left(\frac{1}{k^2}\right) = \frac{h^2}{\log^2(2)}\frac{1}{k} + O\left(\frac{1}{k^2}\right). \qquad (41)$$

Therefore, asymptotically the ratio of the variances $\frac{\text{Var}(\hat{h}_{me})}{\text{Var}(\hat{h}_{gm})} = \frac{6}{\pi^2 \log^2(2)} \approx 1.27$. In other words, asymptotically, our proposed geometric mean estimator is about 27% more accurate than the sample median estimator when $\alpha \to 0+$.

Non-asymptotically, however, the geometric mean estimator can be much more accurate when $k$ is not too large. The moments of the sample median estimator $\hat{T}_{me}$ can be written in an integral form.

$$\mathbf{E}\left(\hat{h}_{me}^s\right) = h^s \int_0^\infty \frac{t^s}{1/\log^s(2)}\left(\exp\left(-\frac{1}{t}\right)\right)^m \left(1 - \exp\left(-\frac{1}{t}\right)\right)^m \exp\left(-\frac{1}{t}\right)\frac{1}{t^2}\frac{(2m+1)!}{(m!)^2}dt$$
$$= h^s \int_0^1 \frac{\log^s(2)}{(-\log(t))^s}(t - t^2)^m \frac{(2m+1)!}{(m!)^2}dt, \qquad (42)$$

based the properties of sample quantiles (see [56, Example 2.9]). For convenience, we only consider $k = 2m + 1$. When $s = 1$ and 2, this integral can actually be expressed as finite (binomial-type) summations [37, 4.267.41, 4.268.5], which, however, are numerically unstable when $m > 12$.

We can compare the estimators in terms of their mean square errors (MSE's). The MSE of $\hat{h}_{me}$ is $\infty$ if $k < 5$ and it is about 3.26 times (which is very considerable) as large as that of $\hat{h}_{gm}$ if $k = 5$. The ratio of their MSE's, converges to about 1.27 as $k \to \infty$, as illustrated in Figure 9.

The above analysis reveals why it is not convenient to study the sample median estimator, even the probability density function is explicitly available.

It is even more difficult to analyze the tail bounds for the sample median estimator, especially if we want the explicit constants. The basic result on tail bounds was from [29], which derived the "pseudo-exponential" tail bounds for the sample quantiles with un-specified constants. Our technique for deriving tail bounds for the geometric mean estimator may be also applicable to the sample median estimator. That is, we can use the Markov moment bound together with the asymptotic properties of the sample median to derive sub-optimal (but asymptotic optimal) tail bounds, if we do not mind bounds in double integral forms. There are "distribution-free" bounds for the moments of order statistics [23], but only when the data have bounded first and second moments.

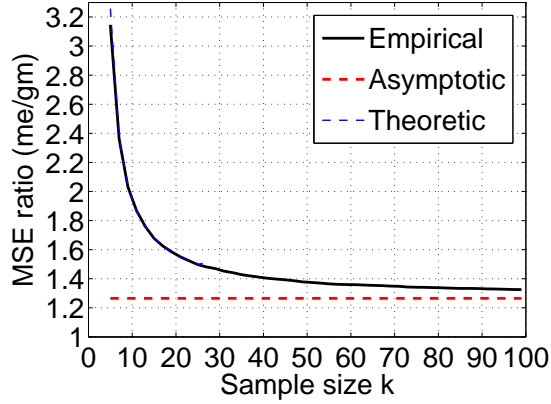

Figure 9: Ratios of the MSE ($\hat{h}_{me}$ v.s. $\hat{h}_{gm}$). The horizontal (dashed) line is the theoretical asymptotic value (i.e., 1.27). The solid curve is obtained from simulations ($5 \times 10^5$ samples for each $k$). The thin dashed curve (for $k \leq 25$), which overlaps with the solid curve, is the theoretical MSE ratios obtained by theoretically integrating (42).

On the other hand, since we can analyze the geometric mean estimator fairly easily, which is also more accurate, there seems to be no need to struggle for the exact moments and bounds for the sample median estimator.

Finally, we will discuss about the maximum likelihood estimator (MLE), which is asymptotically optimal. For stable distributions, MLE is in general very expensive because we will have to numerically evaluate the probability density function, except when $\alpha = 2, 1, 0+$. The technical report [47] discussed about the MLE for $\alpha = 1$. Here, we will discuss the case $\alpha = 0+$, which is in fact particularly simple for the MLE.

Recall as $\alpha \to 0+$, we have $k$ i.i.d. samples $z_j \sim h/E_1$. It is easy to show $E(\frac{1}{z_j}) = \frac{1}{h}$, which implies a straightforward estimator

$$\hat{h}_{mle} = \frac{1}{\frac{1}{k}\sum_{j=1}^{k} \frac{1}{z_j}}. \tag{43}$$

$\hat{h}_{mle}$ is indeed the maximum likelihood estimator. We recommend the bias-corrected version

$$\hat{h}_{mle,c} = \frac{1}{\frac{1}{k}\sum_{j=1}^{k} \frac{1}{z_j}} \left(1 - \frac{1}{k}\right). \tag{44}$$

The moments of $\hat{h}_{mle}$ and $\hat{h}_{mle,c}$ are analyzed in the following lemma.

**Lemma 6** *The first two moments of the maximum likelihood estimator (43) are*

$$E\left(\hat{h}_{mle}\right) = h\left(1 + \frac{1}{k}\right) + O\left(\frac{1}{k^2}\right) \tag{45}$$

$$Var\left(\hat{h}_{mle}\right) = h^2 \left(\frac{1}{k} + \frac{4}{k^2}\right) + O\left(\frac{1}{k^3}\right). \tag{46}$$

*The first two moments of the bias-corrected maximum likelihood estimator (44) are*

$$E\left(\hat{h}_{mle,c}\right) = h + O\left(\frac{1}{k^2}\right) \tag{47}$$

$$Var\left(\hat{h}_{mle,c}\right) = h^2 \left(\frac{1}{k} + \frac{2}{k^2}\right) + O\left(\frac{1}{k^3}\right). \tag{48}$$

Figure 10 verifies the theoretical asymptotic variance formula for $\hat{h}_{mle,c}$.

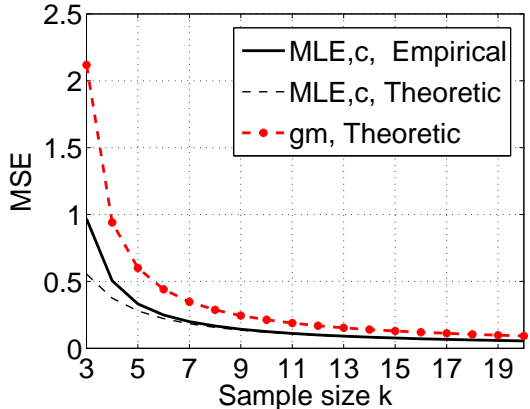

Figure 10: We plot the empirical MSE of the bias-corrected maximum likelihood estimator $\hat{h}_{mle,c}$ ($5\times 10^5$ samples at each $k$) together with the theoretical asymptotic MSE of $\hat{h}_{mle,c}$ up to the third order, i.e., $\left(\frac{1}{k}+\frac{2}{k^2}\right)$ (here we consider $h=1$.) We can see that as soon as $k\geq 7$, our theoretical asymptotic formula is very accurate. The thick dashed curve is the theoretical variance of the unbiased geometric mean estimator $\hat{d}_{gm}$, indicating that $\hat{h}_{mle,c}$ is always more accurate, both asymptotically and non-asymptotically.

The Fisher efficiency of the geometric mean estimator as $\alpha \to 0+$ would be $1/\left(\frac{\pi^2}{6}\right)=0.6079$. For this case, the MLE solution is actually simpler than the geometric mean estimator. Also, notice that if use the regular normal ($l_2$) random projections on the binary-quantized (0/1) data, the estimation variance would be $\frac{2h^2}{k}$ (see [59]), which is about twice as large as Var$\left(\hat{h}_{mle,c}\right)$. This implies that stable random projections with very small $\alpha$ not only provide a solution to approximating the Hamming norms in dynamic settings (i.e., data are subject to frequent additions/subtractions), but also are preferable even in static data.

## 6 Conclusion

Stable random projection is a very useful tool for various applications, such as approximating all pair-wise $l_\alpha$ ($0<\alpha\leq 2$) distances and data stream computations.

In this study, we propose *very sparse stable random projections*, to simply the sampling, to speedup the processing time (i.e., matrix multiplication), and to reduce the storage cost. As shown both theoretically and empirically, it is evident that we can achieve a very significant improvement without hurting the accuracy when $\alpha$ is less than 1 or $\alpha$ is not too larger than 1.

We analyze in detail the estimators based on the geometric mean for recovering the original $l_\alpha$ norms from the projected data. The geometric mean estimators are computationally simple and fairly accurate especially when $\alpha$ is around 1. Moreover, the geometric mean estimators allow us to study the moments precisely and derive practically useful tail bounds in explicit exponential forms, which are otherwise difficult to obtain (e.g.,) from the (commonly used) sample median estimator or the maximum likelihood estimator. An analog of the JL Lemma for dimension reduction in $l_\alpha$ follows immediately from these exponential tail bounds: It suffices to use $k = O\left(\frac{\log(n)}{\epsilon^2}\right)$ projections so that the $l_\alpha$ distance between any pair of data points (among $n$ data points) can be estimated within a $1\pm\epsilon$ factor with high probability, using our proposed geometric mean estimator.

The geometric mean estimators are particularly useful when solving the maximum likelihood estimators is computationally expensive (i.e., when $\alpha \neq 2, 1$, or $0+$). Even when $\alpha = 1$, the maximum likelihood estimator requires solving a high-order polynomial nonlinear equation (see the technical report [47]); and hence it is still considerably more expensive than the geometric mean estimator.

The special case $\alpha \to 0+$ is both practically useful and theoretically interesting. In this case, asymptotically, the geometric mean estimator is 27% more accurate (in terms of variances) than the sample median estimator, and much more non-asymptotically. However, in this case, the (biased-corrected) maximum likelihood estimator can be expressed in a very simple convenient form and is considerably more accurate than the geometric mean estimator, both asymptotically and non-asymptotically. More interestingly, the variance of the maximum likelihood estimator is only about one half of the variance of the regular normal ($l_2$) random projections. Therefore, stable random projections with very small $\alpha$ not only provide a solution to approximating the Hamming norms in dynamic settings (i.e., data are subject to frequent additions/subtractions), but also can be preferable even in static data.

## Acknowledgment

Ping Li thanks Trevor Hastie for his support and many discussions. Ping Li also thanks Kenneth Church for discussions and providing real-world data for numerically verifying the theoretical results. In the course of writing, several theorists offered very constructive comments and suggestions. Their helps are highly appreciated.

Ping Li is partially funded by an NSF Grant DMS0505676.

## A  Proof of Lemma 1

Let $c_i = \frac{g_i}{\left(\sum_{s=1}^{D} |g_s|^\alpha\right)^{\frac{1}{\alpha}}}$. to show the convergence in distribution, it suffices to show the convergence of the characteristic function, i.e.,

$$\log\left(\mathrm{E}\left(\exp\left(\sqrt{-1}t\beta^{-1/\alpha}\sum_{i=1}^{D} c_i z_i\right)\right)\right) \to -\Gamma(1-\alpha)\cos\left(\frac{\pi}{2}\alpha\right)|t|^\alpha \tag{49}$$

By our definition of $z_i$ in (11), we have

$$\mathrm{E}\left(\exp(\sqrt{-1}z_i t)\right) = 1 - \beta + \beta \int_1^\infty \frac{\alpha \cos(xt)}{x^{1+\alpha}} dx$$

$$= 1 - \beta + \alpha\beta\left(\int_0^\infty \frac{\cos(xt)-1}{x^{1+\alpha}}dx - \int_0^1 \frac{\cos(xt)-1}{x^{1+\alpha}}dx + \int_1^\infty \frac{1}{x^{1+\alpha}}dx\right) \tag{50}$$

$\int_1^\infty \frac{1}{x^{1+\alpha}} dx = \frac{1}{\alpha}$. Using the integral formula [37, 3.823, page 484], we obtain

$$\int_0^\infty \frac{\cos(xt)-1}{x^{1+\alpha}}dx = -\frac{1}{\alpha}|t|^\alpha \Gamma(1-\alpha)\cos\left(\frac{\pi}{2}\alpha\right). \tag{51}$$

Also, by the Taylor expansion,

$$\int_0^1 \frac{\cos(xt)-1}{x^{1+\alpha}}dx = -\frac{1}{2}\frac{|t|^2}{2-\alpha} + \frac{1}{4!}\frac{|t|^4}{4-\alpha} + ... \tag{52}$$

Combining the results, we obtain

$$\mathrm{E}\left(\exp(\sqrt{-1}z_i t)\right) = 1 - \beta + \beta\left(1 - |t|^\alpha \Gamma(1-\alpha)\cos\left(\frac{\pi}{2}\alpha\right) + \frac{\alpha}{2}\frac{|t|^2}{2-\alpha} - \frac{\alpha}{4!}\frac{|t|^4}{4-\alpha} + ...\right)$$

$$= 1 - \beta|t|^\alpha \Gamma(1-\alpha)\cos\left(\frac{\pi}{2}\alpha\right) + \frac{\alpha\beta}{2}\frac{|t|^2}{2-\alpha} + ... \tag{53}$$

The above steps are similar to [44]. Once we know $\mathrm{E}\left(\exp(\sqrt{-1}z_i t)\right)$, we can express

$$\log\left(\mathrm{E}\left(\exp\left(\sqrt{-1}t\beta^{-1/\alpha}\sum_{i=1}^{D} c_i z_i\right)\right)\right)$$

$$= \sum_{i=1}^{D} \log\left(1 - |c_i|^\alpha |t|^\alpha \Gamma(1-\alpha)\cos\left(\frac{\pi}{2}\alpha\right) + \frac{\alpha}{2}\frac{|c_i|^2 |t|^2}{\beta^{2/\alpha-1}(2-\alpha)} + ...\right)$$

$$= \sum_{i=1}^{D} \left(-|c_i|^\alpha |t|^\alpha \Gamma(1-\alpha)\cos\left(\frac{\pi}{2}\alpha\right) + \frac{\alpha}{2}\frac{|c_i|^2 |t|^2}{\beta^{2/\alpha-1}(2-\alpha)} - \frac{1}{2}|c_i|^{2\alpha}|t|^{2\alpha}\left(\Gamma(1-\alpha)\cos\left(\frac{\pi}{2}\alpha\right)\right)^2 + ...\right).$$

If $\max_{1 \leq i \leq D}(|c_i|) \to 0$, then

$$\log\left(\mathrm{E}\left(\exp\left(\sqrt{-1}t\beta^{-1/\alpha}\sum_{i=1}^{D}c_i z_i\right)\right)\right) = -|t|^\alpha \Gamma(1-\alpha)\cos\left(\frac{\pi}{2}\alpha\right)\sum_{i=1}^{D}|c_i|^\alpha + \dots$$

$$\to -|t|^\alpha \Gamma(1-\alpha)\cos\left(\frac{\pi}{2}\alpha\right).$$

The rate of convergence is determined by the next higher order term, which could be either the $|t|^2$ term or the $|t|^{2\alpha}$ term, depending on $\alpha$, $\beta$, and the data. The rate of convergence is

$$\begin{cases} O\left(\frac{\sum_{i=1}^{D}|g_i|^2}{\beta^{\alpha/2-1}\left(\sum_{i=1}^{D}|g_i|^\alpha\right)^{2/\alpha}}\right), & \text{if } 1 \leq \alpha < 2 \\ O\left(\max\left\{\frac{\sum_{i=1}^{D}|g_i|^{2\alpha}}{\left(\sum_{i=1}^{D}|g_i|^\alpha\right)^2}, \frac{\sum_{i=1}^{D}|g_i|^2}{\beta^{2/\alpha-1}\left(\sum_{i=1}^{D}|g_i|^\alpha\right)^{2/\alpha}}\right\}\right), & \text{if } 0 < \alpha < 1 \end{cases} \quad (54)$$

This completes the proof of Lemma 1.

## B  Proof of Lemma 2

The estimator defined in (21)

$$\hat{d}_{(\alpha),gm} = \frac{\prod_{j=1}^{k}|x_j|^{1/k}}{\left[\frac{2}{\pi}\Gamma\left(\frac{1}{k}\right)\Gamma\left(1-\frac{1}{k\alpha}\right)\sin\left(\frac{\pi}{2}\frac{1}{k}\right)\right]^k}, \quad (55)$$

is obviously unbiased, i.e., $\mathrm{E}\left(\hat{d}_{(\alpha),gm}\right) = d_{(\alpha)}$, because $x_j$'s are i.i.d. $S(\alpha, d_{(\alpha)})$ with

$$\mathrm{E}\left(|x_j|^{1/k}\right) = d_{(\alpha)}^{1/k}\frac{2}{\pi}\Gamma\left(\frac{1}{k}\right)\Gamma\left(1-\frac{1}{k\alpha}\right)\sin\left(\frac{\pi}{2}\frac{1}{k}\right). \quad (56)$$

The variance of $\hat{d}_{(\alpha),gm}$ is then

$$\mathrm{Var}\left(\hat{d}_{(\alpha),gm}\right) = d_{(\alpha)}^2\left\{\frac{\left[\frac{2}{\pi}\Gamma\left(\frac{2}{k}\right)\Gamma\left(1-\frac{2}{k\alpha}\right)\sin\left(\pi\frac{1}{k}\right)\right]^k}{\left[\frac{2}{\pi}\Gamma\left(\frac{1}{k}\right)\Gamma\left(1-\frac{1}{k\alpha}\right)\sin\left(\frac{\pi}{2}\frac{1}{k}\right)\right]^{2k}} - 1\right\}. \quad (57)$$

It remains to show the asymptotic behaviors of $\hat{d}_{(\alpha),gm}$ and $\mathrm{Var}\left(\hat{d}_{(\alpha),gm}\right)$. First, we will show

$$\left[\frac{2}{\pi}\Gamma\left(\frac{1}{k}\right)\Gamma\left(1-\frac{1}{k\alpha}\right)\sin\left(\frac{\pi}{2}\frac{1}{k}\right)\right]^k \to \exp\left(-\gamma_e\left(1-\frac{1}{\alpha}\right)\right), \text{ as } k \to \infty,$$

where $\gamma_e = 0.577215665...$, is the Euler's constant.

By Euler's reflection formula, $\Gamma(1-z)\Gamma(z) = \frac{\pi}{\sin(\pi z)}$,

$$\left[\frac{2}{\pi}\Gamma\left(\frac{1}{k}\right)\Gamma\left(1-\frac{1}{k\alpha}\right)\sin\left(\frac{\pi}{2}\frac{1}{k}\right)\right]^k = \left[\frac{1}{\alpha}\frac{2\sin\left(\frac{\pi}{2}\frac{1}{k}\right)}{\sin\left(\frac{\pi}{\alpha k}\right)}\right]^k\left[\alpha\frac{\Gamma\left(\frac{1}{k}\right)}{\Gamma\left(\frac{1}{k\alpha}\right)}\right]^k. \quad (58)$$

Because $\left[\frac{1}{\alpha}\frac{2\sin\left(\frac{\pi}{2}\frac{1}{k}\right)}{\sin\left(\frac{\pi}{\alpha k}\right)}\right]^k \to 1$, we only need to study $\left[\alpha\frac{\Gamma\left(\frac{1}{k}\right)}{\Gamma\left(\frac{1}{k\alpha}\right)}\right]^k$.

We use the infinite-product representation of the Gamma function[37, 8.322, page 944],

$$\Gamma(z) = \frac{\exp(-\gamma_e z)}{z}\prod_{s=1}^{\infty}\left(1+\frac{z}{s}\right)^{-1}\exp\left(\frac{z}{s}\right),$$

to obtain

$$\left[\alpha\frac{\Gamma\left(\frac{1}{k}\right)}{\Gamma\left(\frac{1}{k\alpha}\right)}\right]^k = \exp\left(-\gamma_e\left(1-\frac{1}{\alpha}\right)\right)\prod_{s=1}^{\infty}\exp\left(\frac{1}{s}\left(1-\frac{1}{\alpha}\right)\right)\left(1+\frac{1}{ks}\right)^{-k}\left(1+\frac{1}{\alpha ks}\right)^k$$

$$= \exp\left(-\gamma_e\left(1-\frac{1}{\alpha}\right)\right)\exp\left(\sum_{s=1}^{\infty}\left(1-\frac{1}{\alpha}\right)\frac{1}{s} + k\log\left(1+\frac{1}{\alpha ks}\right) - k\log\left(1+\frac{1}{ks}\right)\right)$$

$$= \exp\left(-\gamma_e\left(1-\frac{1}{\alpha}\right)\right)\exp\left(\sum_{s=1}^{\infty}\frac{c_1}{ks^2}+\frac{c_2}{k^2s^3}+\frac{c_3}{k^3s^4}+...\right), \tag{59}$$

where $c_1 = \frac{1}{2}\left(1-\frac{1}{\alpha^2}\right)$, $c_2 = -\frac{1}{3}\left(1-\frac{1}{\alpha^2}\right)$, $c_3 = \frac{1}{4}\left(1-\frac{1}{\alpha^2}\right)$, ..., are the coefficients from the Taylor expansion of the logarithms. Note that $\sum_{s=1}^{\infty}\frac{1}{s^2} = \frac{\pi^2}{6}$, and $\sum_{s=1}^{\infty}\frac{1}{s^{t+1}} < \sum_{s=1}^{\infty}\frac{1}{s^t} < \infty$ for $t \geq 2$. We can then write

$$\sum_{s=1}^{\infty}\frac{c_1}{ks^2}+\frac{c_2}{k^2s^3}+\frac{c_3}{k^3s^4}+... = \sum_{j=1}^{\infty}\frac{c'_j}{k^j} \leq |c'_1|\sum_{j=1}^{\infty}\frac{1}{k^j} = |c'_1|\frac{1/k}{1-1/k} \to 0, \quad \text{as } k \to \infty \tag{60}$$

where $c'_j = c_j\sum_{s=1}^{\infty}\frac{1}{s^{j+1}}$ and obviously $|c'_{j+1}| \leq |c'_j| \leq |c'_1|$. Therefore, we have proved

$$\left[\frac{2}{\pi}\Gamma\left(\frac{1}{k}\right)\Gamma\left(1-\frac{1}{k\alpha}\right)\sin\left(\frac{\pi}{2}\frac{1}{k}\right)\right]^k \to \exp\left(-\gamma_e\left(1-\frac{1}{\alpha}\right)\right), \quad \text{as } k \to \infty. \tag{61}$$

It involves tedious algorithm to check the monotonicity, i.e., $\left[\frac{2}{\pi}\Gamma\left(\frac{1}{k}\right)\Gamma\left(1-\frac{1}{k\alpha}\right)\sin\left(\frac{\pi}{2}\frac{1}{k}\right)\right]^k$ decreases with increasing $k$. Our approach is to also use the infinite-product representation of the sine function [37, 1.431.1, page 44], $\sin(x) = x\prod_{s=1}^{\infty}\left(1-\frac{x^2}{s^2\pi^2}\right)$, so that we could express the whole $\left[\frac{2}{\pi}\Gamma\left(\frac{1}{k}\right)\Gamma\left(1-\frac{1}{k\alpha}\right)\sin\left(\frac{\pi}{2}\frac{1}{k}\right)\right]^k$ as an infinite-product. Then it suffices to show that, for any $s \geq 1$,

$$\log\left(\left(1-\frac{1}{4k^2s^2}\right)^k\left(1-\frac{1}{\alpha^2k^2s^2}\right)^{-k}\left(1+\frac{1}{ks}\right)^{-k}\left(1+\frac{1}{\alpha ks}\right)^k\right)$$

$$= k\log\left(\frac{4\alpha^2k^3s^3 - \alpha^2ks + 4\alpha k^2s^2 - \alpha^2}{4\alpha^2k^3s^3 - 4ks + 4\alpha^2k^2s^2 - 4}\right), \tag{62}$$

is monotonically decreasing. This is a much easier problem. We can take its first two derivatives of (62) and check that the first derivative is monotonically increasing with increasing $k$, reaching 0 when $k = \infty$. Therefore, the first derivative of (62) is negative; and hence we have proved the monotonicity.

Finally, we need to show the asymptotic variance $\hat{d}_{(\alpha),gm}$,

$$\text{Var}\left(\hat{d}_{(\alpha),gm}\right) = d_{(\alpha)}^2\left\{\frac{\left[\frac{2}{\pi}\Gamma\left(\frac{2}{k}\right)\Gamma\left(1-\frac{2}{k\alpha}\right)\sin\left(\pi\frac{1}{k}\right)\right]^k}{\left[\frac{2}{\pi}\Gamma\left(\frac{1}{k}\right)\Gamma\left(1-\frac{1}{k\alpha}\right)\sin\left(\frac{\pi}{2}\frac{1}{k}\right)\right]^{2k}} - 1\right\} = d_{(\alpha)}^2\left(\frac{\pi^2}{6k}\left(\frac{1}{2}+\frac{1}{\alpha^2}\right)+O\left(\frac{1}{k^2}\right)\right).$$

Again, we use Taylor expansions. First, by Euler's reflection formula and some algebra, we obtain

$$\frac{\left[\frac{2}{\pi}\Gamma\left(\frac{2}{k}\right)\Gamma\left(1-\frac{2}{k\alpha}\right)\sin\left(\pi\frac{1}{k}\right)\right]^k}{\left[\frac{2}{\pi}\Gamma\left(\frac{1}{k}\right)\Gamma\left(1-\frac{1}{k\alpha}\right)\sin\left(\frac{\pi}{2}\frac{1}{k}\right)\right]^{2k}} - 1 = \left[\frac{1}{2}\frac{\Gamma\left(\frac{2}{k}\right)\Gamma^2\left(\frac{1}{k\alpha}\right)}{\Gamma^2\left(\frac{1}{k}\right)\Gamma\left(\frac{2}{k\alpha}\right)}\tan\left(\frac{\pi}{k\alpha}\right)\text{ctan}\left(\frac{\pi}{2k}\right)\right]^k - 1. \tag{63}$$

By Taylor expansions,

$$\tan\left(\frac{\pi}{k\alpha}\right)\text{ctan}\left(\frac{\pi}{2k}\right) = \left(\frac{\pi}{k\alpha}+\frac{1}{3}\left(\frac{\pi}{k\alpha}\right)^3+...\right)\left(\frac{2k}{\pi}-\frac{1}{3}\frac{\pi}{2k}+...\right)$$

$$= \frac{2}{\alpha}+\frac{\pi^2}{k^2}\left(\frac{2}{3}\frac{1}{\alpha^3}-\frac{1}{6}\frac{1}{\alpha}\right)+O\left(\frac{1}{k^3}\right) \tag{64}$$

For $\frac{\Gamma\left(\frac{2}{k}\right)\Gamma^2\left(\frac{1}{k\alpha}\right)}{\Gamma^2\left(\frac{1}{k}\right)\Gamma\left(\frac{2}{k\alpha}\right)}$, we again resort to the infinite-product representation of the Gamma function. Some algebra yields

$$\begin{aligned}
\frac{\Gamma\left(\frac{2}{k}\right)\Gamma^2\left(\frac{1}{k\alpha}\right)}{\Gamma^2\left(\frac{1}{k}\right)\Gamma\left(\frac{2}{k\alpha}\right)} &= \alpha \prod_{s=1}^{\infty} \left(1+\frac{2}{ks}\right)^{-1}\left(1+\frac{1}{ks\alpha}\right)^{-2}\left(1+\frac{1}{ks}\right)^{2}\left(1+\frac{2}{ks\alpha}\right) \\
&= \alpha \prod_{s=1}^{\infty} \left(1-\frac{2}{ks}+\frac{4}{k^2s^2}+...\right)\left(1-\frac{2}{ks\alpha}+\frac{3}{k^2s^2\alpha^2}+...\right)\left(1+\frac{2}{ks}+\frac{1}{k^2s^2}\right)\left(1+\frac{2}{ks\alpha}\right) \\
&= \alpha \prod_{s=1}^{\infty} \left(1+\frac{1}{k^2s^2}\left(1-\frac{1}{\alpha^2}\right)+O\left(\frac{1}{k^3}\right)\right) \\
&= \alpha \exp\left(\sum_{s=1}^{\infty} \log\left(1+\frac{1}{k^2s^2}\left(1-\frac{1}{\alpha^2}\right)+O\left(\frac{1}{k^3}\right)\right)\right) \\
&= \alpha \exp\left(\sum_{s=1}^{\infty} \frac{1}{k^2s^2}\left(1-\frac{1}{\alpha^2}\right)+O\left(\frac{1}{k^3}\right)\right) \\
&= \alpha \exp\left(\left(1-\frac{1}{\alpha^2}\right)\frac{1}{k^2}\frac{\pi^2}{6}+O\left(\frac{1}{k^3}\right)\right) \qquad\qquad (\sum_{s=1}^{\infty}\frac{1}{s^2}=\frac{\pi^2}{6}) \\
&= \alpha\left(1+\left(1-\frac{1}{\alpha^2}\right)\frac{1}{k^2}\frac{\pi^2}{6}+O\left(\frac{1}{k^3}\right)\right) \qquad\qquad (65)
\end{aligned}$$

Therefore,

$$\begin{aligned}
&\left[\frac{1}{2}\frac{\Gamma\left(\frac{2}{k}\right)\Gamma^2\left(\frac{1}{k\alpha}\right)}{\Gamma^2\left(\frac{1}{k}\right)\Gamma\left(\frac{2}{k\alpha}\right)}\tan\left(\frac{\pi}{k\alpha}\right)\text{ctan}\left(\frac{\pi}{2k}\right)\right]^k - 1 \\
&= \left[\left(1+\left(1-\frac{1}{\alpha^2}\right)\frac{1}{k^2}\frac{\pi^2}{6}+O\left(\frac{1}{k^3}\right)\right)\left(1+\frac{\pi^2}{k^2}\left(\frac{1}{3}\frac{1}{\alpha^2}-\frac{1}{12}\right)+O\left(\frac{1}{k^3}\right)\right)\right]^k - 1 \\
&= \left[1+\frac{\pi^2}{6k^2}\left(\frac{1}{2}+\frac{1}{\alpha^2}\right)+O\left(\frac{1}{k^3}\right)\right]^k - 1 \\
&= \frac{\pi^2}{6k}\left(\frac{1}{2}+\frac{1}{\alpha^2}\right)+O\left(\frac{1}{k^2}\right), \qquad\qquad (66)
\end{aligned}$$

and this completes of the proof of Lemma 2.

## C  Proof of Lemma 3

We first show the right tail bound

$$\mathbf{Pr}\left(\hat{d}_{(\alpha),gm}-d_{(\alpha)} > \epsilon d_{(\alpha)}\right) \leq \exp\left(-k\frac{\epsilon^2}{M_{\alpha,\epsilon}}\right), \quad \epsilon > 0 \ \ k > \frac{1}{\alpha}$$

where

$$\frac{1}{M_{R,\alpha,\epsilon}} = \frac{1}{c_\alpha \epsilon}\log(1+\epsilon)-\frac{1}{\epsilon^2}\log\left(\frac{2}{\pi}\Gamma\left(1-\frac{\epsilon}{c_\alpha\alpha}\right)\Gamma\left(\frac{\epsilon}{c_\alpha}\right)\sin\left(\frac{\pi}{2}\frac{\epsilon}{c_\alpha}\right)\right)-\frac{1}{c_\alpha\epsilon}\gamma_e\left(1-\frac{1}{\alpha}\right)$$

$$c_\alpha = \frac{\pi^2}{6}\left(\frac{1}{2}+\frac{1}{\alpha^2}\right)$$

First, we note that $\hat{d}_{(\alpha),gm} = \frac{\prod_{j=1}^{k}|x_j|^{1/k}}{\left[\frac{2}{\pi}\Gamma\left(\frac{1}{k}\right)\Gamma\left(1-\frac{1}{k\alpha}\right)\sin\left(\frac{\pi}{2}\frac{1}{k}\right)\right]^k}$ only has moments less than $k$. Therefore, we have to use the Markov moment bound instead of the more often used Chernoff bound[17]. As a matter of fact, for positive random variables, e.g., $\hat{d}_{(\alpha),gm}$, the moment bound is always sharper than the Chernoff bound[54, 50].

Applying the moment bound,

$$\mathbf{Pr}\left(\hat{d}_{(\alpha),gm}-d_{(\alpha)} > \epsilon d_{(\alpha)}\right) \leq \frac{\mathbf{E}\left(\hat{d}_{(\alpha),gm}\right)^t}{(1+\epsilon)^t d_{(\alpha)}^t}, \quad t > 0 \qquad\qquad (67)$$

Usually, we will then have to find the optimal $t$ that minimizes the upper bound. This is difficult in our case, because the optimization will involve the Gamma functions. We avoid this problem by a trick.

We know how to choose the optimal $t$ for a gamma random variable[54, Example 3.3]. $\hat{d}_{(\alpha),gm}$ in fact resembles a gamma. As $k$ increases, $\hat{d}_{(\alpha),gm}$ approaches a normal asymptotically, but it can be better characterized by an asymptotically equivalent gamma (i.e., with the same mean and variance), because $\hat{d}_{(\alpha),gm}$ is positive and has non-zero odd central moments.

Suppose $z$ is gamma with mean $\mu$ and variance $\sigma^2$, then $\mathbf{Pr}\left(z > (1+\epsilon)\mu\right) < \frac{\mathbf{E}(z^t)}{(1+\epsilon)^t \mu^t}$, which is minimized at $t = 1 + \lfloor \epsilon \frac{\mu^2}{\sigma^2} \rfloor$ when $t$ is restricted to be integers. This can be inferred directly from [54, Example 3.3]. Since the moment bound is true for any $t > 0$, we could for convenience choose $t = \epsilon \frac{\mu^2}{\sigma^2}$.

Now, if we assume that $\hat{d}_{(\alpha),gm}$ is a gamma random variable with mean $d_{(\alpha)}$ and variance $d_{(\alpha)}^2 \frac{\pi^2}{6k} \left(\frac{1}{2} + \frac{1}{\alpha^2}\right)$ (for convenience, we consider the asymptotic variance), then we can infer the "optimal" $t$ value to be

$$t = \frac{6k\epsilon}{\pi^2 \left(\frac{1}{2} + \frac{1}{\alpha^2}\right)} = \frac{\epsilon}{c_\alpha} k, \quad \text{where} \quad c_\alpha = \frac{\pi^2}{6}\left(\frac{1}{2} + \frac{1}{\alpha^2}\right), \tag{68}$$

which is, of course, indeed only sub-optimal because $\hat{d}_{(\alpha),gm}$ is not a gamma non-asymptotically. However, because the moment bound (67) holds for any $t > 0$, we can plug in $t = \frac{\epsilon}{c_\alpha} k$ and obtain

$$\mathbf{Pr}\left(\hat{d}_{(\alpha),gm} - d_{(\alpha)} > \epsilon d_{(\alpha)}\right) \leq \left[\frac{\frac{2}{\pi}\Gamma\left(1 - \frac{\epsilon}{c_\alpha \alpha}\right)\Gamma\left(\frac{\epsilon}{c_\alpha}\right)\sin\left(\frac{\pi}{2}\frac{\epsilon}{c_\alpha}\right)}{\left(\frac{2}{\pi}\Gamma\left(1 - \frac{1}{k\alpha}\right)\Gamma\left(\frac{1}{k}\right)\sin\left(\frac{\pi}{2}\frac{1}{k}\right)\right)^{\frac{\epsilon}{c_\alpha}k}(1+\epsilon)^{\frac{\epsilon}{c_\alpha}}}\right]^k$$

$$\leq \left[\frac{\frac{2}{\pi}\Gamma\left(1 - \frac{\epsilon}{c_\alpha \alpha}\right)\Gamma\left(\frac{\epsilon}{c_\alpha}\right)\sin\left(\frac{\pi}{2}\frac{\epsilon}{c_\alpha}\right)}{\exp\left(-\frac{\epsilon}{c_\alpha}\gamma_e\left(1 - \frac{1}{\alpha}\right)\right)(1+\epsilon)^{\frac{\epsilon}{c_\alpha}}}\right]^k = \exp\left(-k\frac{\epsilon^2}{M_{R,\alpha,\epsilon}}\right),$$

where

$$\frac{1}{M_{R,\alpha,\epsilon}} = \frac{1}{c_\alpha \epsilon}\log(1+\epsilon) - \frac{1}{\epsilon^2}\log\left(\frac{2}{\pi}\Gamma\left(1 - \frac{\epsilon}{c_\alpha \alpha}\right)\Gamma\left(\frac{\epsilon}{c_\alpha}\right)\sin\left(\frac{\pi}{2}\frac{\epsilon}{c_\alpha}\right)\right) - \frac{1}{c_\alpha \epsilon}\gamma_e\left(1 - \frac{1}{\alpha}\right). \tag{69}$$

In the above derivation, we have used the result from Lemma 2, which says, if $k > \frac{1}{\alpha}$, then

$$\left[\frac{2}{\pi}\Gamma\left(\frac{1}{k}\right)\Gamma\left(1 - \frac{1}{k\alpha}\right)\sin\left(\frac{\pi}{2}\frac{1}{k}\right)\right]^k \geq \exp\left(-\gamma_e\left(1 - \frac{1}{\alpha}\right)\right).$$

Next, we will show the left tail bound

$$\mathbf{Pr}\left(\hat{d}_{(\alpha),gm} - d_{(\alpha)} < -\epsilon d_{(\alpha)}\right) \leq \exp\left(-k\frac{\epsilon^2}{M_{L,\alpha,\epsilon,k_0}}\right), \quad 0 < \epsilon \leq 1, k > k_0 > \frac{1}{\alpha}, \tag{70}$$

where

$$\frac{1}{M_{L,\alpha,\epsilon,k_0}} = -\frac{1}{c_\alpha \epsilon}\log(1-\epsilon) - \frac{1}{\epsilon^2}\log\left(-\frac{2}{\pi}\Gamma\left(1 + \frac{\epsilon}{c_\alpha \alpha}\right)\Gamma\left(-\frac{\epsilon}{c_\alpha}\right)\sin\left(\frac{\pi}{2}\frac{\epsilon}{c_\alpha}\right)\right)$$

$$- \frac{1}{c_\alpha \epsilon}k_0 \log\left(\left[\frac{2}{\pi}\Gamma\left(\frac{1}{k_0}\right)\Gamma\left(1 - \frac{1}{k_0 \alpha}\right)\sin\left(\frac{\pi}{2}\frac{1}{k_0}\right)\right]\right). \tag{71}$$

For any $0 \leq t < k$,

$$\mathbf{Pr}\left(\hat{d}_{(\alpha),gm} \leq (1-\epsilon)d_{(\alpha)}\right)$$

$$=\mathbf{Pr}\left(\prod_{j=1}^{k}|x_j|^{-t/k} \geq \left((1-\epsilon)d_{(\alpha)}\left[\frac{2}{\pi}\Gamma\left(\frac{1}{k}\right)\Gamma\left(1-\frac{1}{k\alpha}\right)\sin\left(\frac{\pi}{2}\frac{1}{k}\right)\right]^k\right)^{-t}\right)$$

$$\leq \mathbf{E}\left(\prod_{j=1}^{k}|x_j|^{-t/k}\right)\left((1-\epsilon)d_{(\alpha)}\left[\frac{2}{\pi}\Gamma\left(\frac{1}{k}\right)\Gamma\left(1-\frac{1}{k\alpha}\right)\sin\left(\frac{\pi}{2}\frac{1}{k}\right)\right]^k\right)^{t}$$

$$=\left[-\frac{2}{\pi}\Gamma\left(-\frac{t}{k}\right)\Gamma\left(1+\frac{t}{k\alpha}\right)\sin\left(\frac{\pi}{2}\frac{t}{k}\right)\right]^k\left((1-\epsilon)\left[\frac{2}{\pi}\Gamma\left(\frac{1}{k}\right)\Gamma\left(1-\frac{1}{k\alpha}\right)\sin\left(\frac{\pi}{2}\frac{1}{k}\right)\right]^k\right)^{t}. \quad (72)$$

Again, we need to find the optimal $t$. To avoid the difficulty, we "borrow" the sub-optimal $t = \frac{\epsilon}{c_\alpha}k$, from the right tail bound. The rationale is that, if $\hat{d}_{(\alpha),gm}$ were symmetric about $d_{(\alpha)}$, then the optimal $t$ values will be the same for both tails. In our case, $\hat{d}_{(\alpha),gm}$ is positive skewed with the skewness decreases with increasing $k$. Using $t = \frac{\epsilon}{c_\alpha}k$ will be conservative but should not deviate too much from the true optimum when $k$ is not too small.

Plugging in $t = \frac{\epsilon}{c_\alpha}k$, we obtain, after some algebra,

$$\mathbf{Pr}\left(\hat{d}_{(\alpha),gm} - d_{(\alpha)} < -\epsilon d_{(\alpha)}\right) \leq \exp\left(-k\frac{\epsilon^2}{M_{L,\alpha,\epsilon,k}}\right), \quad 0 < \epsilon \leq 1, \quad (73)$$

where

$$\frac{1}{M_{L,\alpha,\epsilon,k}} = -\frac{1}{c_\alpha\epsilon}\log(1-\epsilon) - \frac{1}{\epsilon^2}\log\left(-\frac{2}{\pi}\Gamma\left(1+\frac{\epsilon}{c_\alpha\alpha}\right)\Gamma\left(-\frac{\epsilon}{c_\alpha}\right)\sin\left(\frac{\pi}{2}\frac{\epsilon}{c_\alpha}\right)\right)$$

$$-\frac{1}{c_\alpha\epsilon}k\log\left(\left[\frac{2}{\pi}\Gamma\left(\frac{1}{k}\right)\Gamma\left(1-\frac{1}{k\alpha}\right)\sin\left(\frac{\pi}{2}\frac{1}{k}\right)\right]\right). \quad (74)$$

We have proved in Lemma 2 that $k\log\left(\left[\frac{2}{\pi}\Gamma\left(\frac{1}{k}\right)\Gamma\left(1-\frac{1}{k\alpha}\right)\sin\left(\frac{\pi}{2}\frac{1}{k}\right)\right]\right) \to -\gamma_e\left(1-\frac{1}{\alpha}\right)$ monotonically from above if $k > \frac{1}{\alpha}$. Therefore, if $k > k_0$, then $\frac{1}{M_{L,\alpha,\epsilon,k}} < \frac{1}{M_{L,\alpha,\epsilon,k_0}}$; and hence we have completed the proof.

## D  Proof of Lemma 6

Assume $z \sim h/E_1$, where $E_1$ stands for the exponential distribution with mean 1. The log likelihood, $l(z;h)$, and first three derivatives (w.r.t. $h$) are

$$l(z;h) = -\frac{h}{z} + \log(h) - 2\log(z) \quad (75)$$

$$l'(h) = \frac{1}{h} - \frac{1}{z} \quad (76)$$

$$l''(h) = -\frac{1}{h^2} \quad (77)$$

$$l'''(d) = \frac{2}{h^3}. \quad (78)$$

The MLE $\hat{h}_{mle}$ is asymptotically normal with mean $h$ and variance $\frac{1}{k\mathrm{I}(h)}$, where $\mathrm{I}(h)$, the expected Fisher Information, is

$$\mathrm{I} = \mathrm{I}(h) = \mathbf{E}\left(-l''(h)\right) = \frac{1}{h^2}. \quad (79)$$

General formulas for the bias and higher moments of the MLE are available in [8, 57]:

$$\mathbf{E}\left(\hat{h}_{mle}\right) = h - \frac{[12]}{2k\mathrm{I}^2} + O\left(\frac{1}{k^2}\right) \quad (80)$$

$$\mathrm{Var}\left(\hat{h}_{mle}\right) = \frac{1}{k\mathrm{I}} + \frac{1}{k^2}\left(-\frac{1}{\mathrm{I}} + \frac{[1^4] - [1^22] - [13]}{\mathrm{I}^3} + \frac{3.5[12]^2 - [1^3]^2}{\mathrm{I}^4}\right) + O\left(\frac{1}{k^3}\right) \quad (81)$$

where, after re-formatting,

$$[12] = E(l')^3 + E(l'l''), \qquad [1^4] = E(l')^4, \qquad [1^2 2] = E(l''(l')^2) + E(l')^4,$$
$$[13] = E(l')^4 + 3E(l''(l')^2) + E(l'l'''), \qquad [1^3] = E(l')^3. \tag{82}$$

Note that, for any integer $m > 0$,

$$E\left(\frac{h}{z}\right)^m = \int_0^\infty \left(\frac{h}{z}\right)^m \exp\left(-\frac{h}{z}\right) \frac{h}{z^2} dz = \int_0^\infty s^m \exp(-s) ds = m \int_0^\infty s^{m-1} \exp(-s) ds = m!, \tag{83}$$

from which it follows that

$$E(l')^3 = -\frac{2}{h^3}, \quad E(l'l'') = 0, \quad E(l')^4 = \frac{9}{h^4}, \quad E(l''(l')^2) = -\frac{1}{h^4}, \quad E(l'l''') = 0. \tag{84}$$

Hence

$$[12] = -\frac{2}{h^3}, \quad [1^4] = \frac{9}{h^4}, \quad [1^2 2] = \frac{8}{h^4}, \quad [13] = \frac{6}{h^4}, \quad [1^3] = -\frac{2}{h^3}. \tag{85}$$

Thus, we obtain

$$E\left(\hat{h}_{mle}\right) = h + \frac{h}{k} + O\left(\frac{1}{k^2}\right) \tag{86}$$

$$\text{Var}\left(\hat{h}_{mle}\right) = \frac{h^2}{k} + \frac{4h^2}{k^2} + O\left(\frac{1}{k^3}\right) \tag{87}$$

Because $\hat{h}_{mle}$ has $O\left(\frac{1}{k}\right)$ bias, we recommend the bias-corrected estimator

$$\hat{h}_{mle,c} = \hat{h}_{mle}\left(1 - \frac{1}{k}\right), \tag{88}$$

whose first two moments are

$$E\left(\hat{h}_{mle,c}\right) = h + O\left(\frac{1}{k^2}\right) \tag{89}$$

$$\text{Var}\left(\hat{h}_{mle,c}\right) = \left(1 - \frac{1}{k}\right)^2 \left(\frac{h^2}{k} + \frac{4h^2}{k^2}\right) + O\left(\frac{1}{k^3}\right) = \frac{h^2}{k} + \frac{2h^2}{k^2} + O\left(\frac{1}{k^3}\right). \tag{90}$$

This completes the proof of Lemma 6.